\documentclass[aps,showpacs,floatfix,superscriptaddress,onecolumn,showkeys]{revtex4-1}

\usepackage{graphicx}
\usepackage{amsmath}
\usepackage{amssymb}
\usepackage{dsfont}

\def\vec#1{{\boldsymbol{#1}}}
\def\<{\langle}
\def\>{\rangle}

\def\set#1{{\sf #1}}

\def\Intgrs{\mathbb Z}

\def\Tr{\operatorname{Tr}}

\begin{document}
\title{Discrete time quantum walks on percolation graphs}

\author{B\'alint Koll\'ar}
\affiliation{Wigner RCP, SZFKI, Konkoly-Thege M. u. 29-33, H-1121 Budapest, Hungary}

\author{Jaroslav Novotny}
\affiliation{Department of Physics, Faculty of Nuclear Sciences and Physical Engineering, Czech Technical University in Prague, B\v
rehov\'a 7, 115 19 Praha 1 - Star\'e M\v{e}sto, Czech Republic}

\author{Tam\'as Kiss}
\affiliation{Wigner RCP, SZFKI, Konkoly-Thege M. u. 29-33, H-1121 Budapest, Hungary}

\author{Igor Jex}
\affiliation{Department of Physics, Faculty of Nuclear Sciences and Physical Engineering, Czech Technical University in Prague, B\v
rehov\'a 7, 115 19 Praha 1 - Star\'e M\v{e}sto, Czech Republic}



\date{\today}

\begin{abstract}
Randomly breaking connections in a graph alters its transport properties, a model used to describe percolation. In the case of quantum walks, dynamic percolation graphs represent a special type of imperfections, where the connections appear and disappear randomly in each step during the time evolution. The resulting open system dynamics is hard to treat numerically in general. We shortly review the literature on this problem. We then present our method to solve the evolution on finite percolation graphs in the long time limit, applying the asymptotic methods concerning random unitary maps. We work out the case of one dimensional chains in detail and provide a concrete, step by step numerical example in order to give more insight into the possible asymptotic behavior. The results about the case of the two-dimensional integer lattice are summarized, focusing on the Grover type coin operator.
\end{abstract}

\maketitle
%

%%%%%%%%%%%%%%%%%%%%%%%%%%%%%
\section{Introduction}
%%%%%%%%%%%%%%%%%%%%%%%%%%%%%
Quantum walks (QWs) represent a new paradigm in the studies describing quantum excitations propagating in a medium. First, quantum walks have been introduced \cite{Aharonov1993,Meyer1996} as the quantum generalization of classical random walks. Since then, they gained considerable attention due to their significant potential: Quantum walks can model transport \cite{Mulken2007}, show special wave phenomena \cite{Inui2004,Stefanak2008a,Shikano2010,Paparo2012,Asboth2013}, and, can be used as universal quantum computational primitives \cite{Childs2009,Lovett2010} in quantum information theory. For reviews, see \cite{Kempe2003,Konno2008,Santha2008,Elias2012}. Recently, several quantum walk based experiments have been performed on various physical systems \cite{Karski2009,Schmitz2009,Zahringer2010,OBrien2010,Broome2010,Schreiber2010,Schreiber2011,Schreiber2012,Sansoni2012,Torres2012,Silberberg2012,Alberti2013,Crespi2013,Thompson2013}. These experimental realizations further motivate the theoretical study of  quantum walks. 

The quantum generalization of random walks has several flavors. The first model is the ``discrete time quantum walk" (or ``coined quantum walk")  \cite{Aharonov1993}, where the rules of movement are described with a unitary operator --- resulting in a single step under unit time. The unitary time evolution operator encodes the structure of the graph on which the walk takes place.  Dynamics of the walk is simply given by the repeated application of the unitary evolution operator on the initial state of the system. This model bears a non-trivial addition: the so-called coin, which refers to an internal degree of freedom of the particle, controlling directions of the possible displacement on a lattice or edges towards neighboring graph vertices in a more general case..
There exists a similar discrete time model, which is based on the description of photons walking in an array of beam-splitters \cite{Hillery2003}. Due to its origins,  this model is called ``scattering quantum walk".
The next model is the ``continuous quantum walk" \cite{Farhi1998}, where the movement of the particle is simply described by a Hamiltonian. Thus, the Hamiltonian encodes the structure of the underlying graph.
The last approach is to generalize a classical Markov chain to the quantum domain, resulting a quantum Markov chain \cite{Szegedy2004}. This non-trivial generalization is called ``Szegedy's quantum walk". In this paper we will focus mainly on the first, discrete time quantum walk model.

All  quantum walk models are examples of deterministic (unitary) evolution. However, when the walk is implemented on a physical system it will necessarily suffer from imperfections. There are many possible source of imperfections  \cite{Kendon2007,Sinayskiy2013,Konno2013,Ahlbrecht2011} and some of them might even lead to new physical phenomena. Among such, we can name the exponential localization \cite{Schreiber2011,Crespi2013}, which happens during the transition from the conducting state of a material to the insulator regime. In this case, the coupling between particular positions have a random, but static variation, which might cause exponential localization in the wave function of the walking particle. An extreme limit of such static randomness is the case of broken links or percolation graphs \cite{Percolation,Percolation2,Kendon2010}: in this limit the underlying graph of the system is changed. However, in some cases the source of randomness can change the system dynamically, throughout the time evolution. An example of such behavior is the so-called dynamical percolation \cite{DynamicalPercolation,Romanelli2005,Romanelli2011,Oliveira2006,Abal2008,Annabestani2010,Marquezino2008,Santos2014,Kollar2012,Kollar2014}: Here, the connectivity of the underlying graph can change from time to time. In this paper we focus on such a model of imperfections.

The paper is organized as follows. First, we briefly review the results of the  literature about quantum walks on percolation graphs. Section \ref{Sec:Defs} is devoted to give the basic definitions of the open system we study. In the next section we present the analytical tools which help us to determine the asymptotic dynamics of the model. In Section \ref{1dqws} we give the full explicit solution for one-dimensional quantum walks, which complement our results published earlier. Following that, we give a case study which illustrates our methods and results. In Section \ref{Sec:2DQWS} we briefly review our results on two-dimensional percolation lattices. Finally, we discuss the results presented in the paper.

%%%%%%%%%%%%%%%%%%%%%%%%%%%%%
\section{Brief overview of the results on percolation quantum walks}
%%%%%%%%%%%%%%%%%%%%%%%%%%%%%
\label{Sec:Reviews}

In this section we shortly review the literature on quantum walks on percolation graphs. Most of the work has been done numerically by simulating the evolution of the system. Brute force methods have rather limited efficiency here, since both the generation of random graphs and the simulation of a quantum evolution are increasingly hard with the growing size of the system.

The earliest work on quantum walks on graphs with broken links is, to our knowledge, by Romanelli \textit{et al.} \cite{Romanelli2005}. They investigated the one-dimensional discrete time Hadamard quantum walk on the line under the decoherence effect of dynamical percolation (broken links). Their model has a single parameter corresponding to errors of the graph: $p$, which is the probability that a link between any two adjacent sites is missing under unit time step. In this model the time evolution is always unitary, thus the effect of broken links is considered as unitary noise. The question the authors have addressed was whether the evolution on such a dynamically changing graph affects the variance (spreading) of the quantum walk. They found numerically that a transition between the two-peaked quantum (ballistically spreading) and classical (diffusively spreading) Gaussian distributions happens around the critical time
$t_c = \frac{1}{p\sqrt{2}}$.
A simple argument behind this result can be given to help the understanding of the previous observation: At the early times the wave-function is confined to a small region. Consequently, it is not disturbed by the dynamical percolation of the graph, thus, the walk can spread ballistically. At time $t$, the walk covers $t / \sqrt{2}$ sites, and about $pt / \sqrt{2}$ links are broken on that area. As the number of broken links grows to the order of $1$, the disturbance becomes relevant, and the quantum walk will lose its quadratic speedup, reverting to the diffusive spreading. The diffusion coefficient is estimated to be $D \simeq 0.4 \frac{(1-p)}{p}$. The authors also determined a critical value for $p \approx 0.44$, when the diffusion coefficient is $1/2$, corresponding to that of the classical unbiased random walk.

The global chirality distribution (GCD) \cite{Romanelli2011} and the the coin (chirality) reduced density matrix \cite{HinajerosPrePrint} of the one-dimensional quantum walk model with broken links was investigated.  The GCD can be described by a Markovian process and the master equation can be obtained analytically. In the the case when only a half-line is affected by the possible breakage, the process is not Markovian. The wave-function on the decoherence free half-line will never be influenced by the decoherence of the other half-line, \textit{i. e.} the wave-function keeps some information from its initial state, hence the system is non-Markovian. The chiral density matrix has a well defined asymptotic limit, it exhibits non-Markovian behavior.

The two-dimensional extension of the above dynamical percolation model was first considered by Oliveira \textit{et al.} \cite{Oliveira2006}. The two-dimensional Hadamard, Grover and Fourier walks were studied in terms of the diffusion coefficient. The authors showed that if the percolation probabilities can be tuned independently on the diagonal line, the walker can become confined to that one-dimensional region. This confinement can lead to increased coherence (and thus, ballistic spreading). Hence, at the extreme cases of low $p \ll 1$ and high $p \approx 1$ the system behaves as a ballistically spreading coherent wave, whereas in the regime between the decoherence is significant, the walk is diffusive. The diffusion coefficient for a Hadamard coin was proposed to be proportional to $\frac{(1-p)}{p^2}$ in \cite{Ampadu2012}.

Abal \textit{et al.} \cite{Abal2008}  investigated the one-dimensional infinite line with broken links using a single-parameter coin class. They treated the model by introducing a translationally invariant type of the dynamical percolation. With probability $p^2$ the walker cannot move, with probability $(1-p)p$ the walker is not displaced to the left (or to the right). Finally, with probability $(1-p)^2$, the walker is free to move (performs an undisturbed step). The authors have successfully determined  the dependence of the diffusion coefficient on the parameter of the coin numerically.
In the article of Annabestani \textit{et al.} \cite{Annabestani2010} the authors state that  the $D=1/2$ diffusion coefficient  (which is the same as for an unbiased classical random walk) is achieved at the critical probability $p=0.417$. Below the critical threshold (when the movement is less restricted), the diffusion coefficient of the quantum walk is higher than the classical walk, and above it is higher.

In the work of Leung \textit{et al.} \cite{Kendon2010} the one-dimensional lattice with dynamically broken links is investigated considering the statistical mixture of the unitary trajectories. Their results about the one-dimensional system agree with the above reviewed results. Furthermore, they claimed that the transition from ballistic to diffusive motion happens slowly in certain cases, thus the quantum speedup might be still exploited. For larger systems they found that the spreading is diffusive, however, the pre-factor of  the spreading of the quantum walk can be still higher than its classical counterpart, \textit{i. e.} its motion is diffusive but faster.
In the same work the Grover walk on a two-dimensional Cartesian lattice with static  bond and the site percolation was analyzed. The authors numerically determined the spreading (variance) of the system. Their results show, that below the critical bond (site) percolation threshold $p \approx 0.5$ $(p \approx 0.6)$ the quantum walk --- like a classical walk --- can not spread. However, above the threshold the spreading of the system shows a fractional scaling, \textit{i. e.} a sub-diffusive motion. In the limit of small number of broken links the quantum walk surpasses the classical diffusive spreading and exhibits sub-ballistic fractional spreading.
In a related article Lovett \textit{et al.} \cite{Lovett2011preprint} numerically investigated percolation graphs as a factor affecting the efficiency of a quantum walk based search on two- and three-dimensional lattices. They found that below the percolation threshold the search fails naturally, since with high probability the graph is not connected. Consequently, the probability amplitude can not concentrate (interfere constructively) on the marked vertices. However, just above the critical percolation threshold the authors found that the walk exceeds the speed of a classical search $\mathcal{O}(N)$. The reasoning behind this effect is that in this case in the two-dimensional percolation graph the remaining structure resembles a one-dimensional graph. Moreover, further above this regime the speed of search rapidly converges to the quantum value $\mathcal{O}(\sqrt{N})$. Surprisingly, this quantum limit is reached around $p=0.7$ (\textit{i. e.} the probability for any edge to be broken).

Chandrashekar \textit{et al.} \cite{ChandrashekarPrePrint} studied the split-step (or directed graph) quantum walk on two-dimensional percolation lattices (the Cartesian and the honeycomb) as a model of transport. With extensive numerical simulations they determined the percolation threshold on finite systems, \textit{i. e.} the critical probability when the walker cannot travel through the lattice. They found that as the size of the underlying lattice increases, the percolation threshold tends to one. The authors claim that the high threshold is caused by the fact, that most part of the wave function suffers localization due to the missing edges --- hence, just only a small number of disconnected edges can possibly disturb the transport on a large system.
In a related article \cite{Chandrashekar2014} the authors numerically investigated a model for an optical network based experiment, where the components might suffer from imperfections. Such imperfections of the physical devices --- reflections and backscattering --- influence the motion of the photon propagating in the optical network: It may percolate through, become localized in the network, or, suffer backscattering. The numerical results of the paper shows, that even a relative small number of defective paths $\approx 10\% $ can break down the probability of the photon percolating through. The authors implied, that additional strategies might be required for practical systems since such high sensitivity to defects.

In the work by Motes \textit{et al.} \cite{MotesPrePrint} the authors investigated a similar property, \textit{i. e.} the escaping probability of the quantum walk on a small two-dimensional site percolation Cartesian lattice. Their numerical studies show that below the critical percolation threshold $\approx 0.6$ the particle has a near zero escape probability. On the other hand, above the percolation threshold (when the number of missing connections are less) the probability of escape jumps to a finite value both in the classical and quantum case. This jump is much more prominent in the quantum case, due to the faster spreading.

Marquezino \textit{et al.} \cite{Marquezino2008} dealt with the discrete time quantum walk on the $n$-dimensional hypercube. The average limiting distribution 
$$ 
\pi(x) = \lim_{T \rightarrow \infty} \frac{1}{T} \sum_{t=0}^{T-1} P(x,t)\,,
$$
was considered in their work, where $P(x,t)$ is the position distribution of the walk at time $t$. The authors employed the dynamical percolation (broken links) as a type of unitary noise --- \textit{i. e.} no averaging over different percolation lattices was performed. In the unperturbed ($p=0$, no broken links) Grover operator driven case the average limiting distribution is not necessarily uniform, depending on the initial state. However, in the percolation case even a small noise will cause the system to reach the uniform limiting distribution. The authors used the mixing time to characterize the speed of convergence. They found that it depends on the probability $p$, and the numerical results imply, that the fastest mixing happens around $p_c \approx 0.1$. Consequently, a small decoherence can aid the mixing procedure.

The article of Santos \textit{et al.} \cite{Santos2014} focused on another discrete time model of the quantum walk: Szegedy's quantum walk \cite{Szegedy2004}. The authors studied the decoherence caused by dynamical percolation. This scenario is handled by  averaging the unitary time evolution operators over all possible sequences --- \textit{i. e.} instances of percolation hypercubes.  Their results imply that the formula of the quantum hitting time under such decoherence obtains a new term, which depends linearly on $p$ (probability of an edge to be broken under unit time). The authors state, that in this way the quantum speed up of the hitting time is preserved for some range of $p$. 

In continuous time quantum walks (CTQWs), the static percolation (where the disorder does not change through the evolution) was considered 
in the works of M\"ulken \textit{et al.} \cite{Mulken2010,Mulken2011} and Anishchenko \textit{et al.} \cite{Anishchenko2012}. These static percolation model is usually referred to as ``statistical networks".

The case of dynamical percolation with CTQWs has been studied by Dar\'azs \textit{et al.} \cite{Darazs2013}. The authors showed, that the dynamical percolation act as a rescaling of time evolution, when the changes occur with a high enough frequency. The return probability was also investigated in detail. It is shown, that although the system suffers a strong decoherence due to the rapid changes of the underlying lattice, the return time still shows an oscillatory behavior, which is a characteristic property of the quantum evolution. In contrast, classical systems show an exponentially decaying return probability.

In our earlier papers \cite{Kollar2012,Kollar2014} we studied the asymptotic dynamics of the discrete time quantum walk model on finite percolation graphs. We used the dynamical percolation model, while also including the statistical averaging over the random realizations. Such model can be viewed as a special type of open system dynamics. We have given an elaborate method to solve the asymptotics of the model, and obtained close form solutions on the studied cases.
The following sections are dedicated to review and extend our results and methods.
In the next section we give the definition of the model.

%%%%%%%%%%%%%%%%%%%%%%%%%%%%%
\section{Definitions}
%%%%%%%%%%%%%%%%%%%%%%%%%%%%%
\label{Sec:Defs}

A unitary discrete time (or coined) quantum walk on a graph moves similarly as a classical random walker would: from a vertex it might hop to its neighboring vertices. Due to the inherent quantum nature of the particle (walker), this hopping is done in a coherent (deterministic) manner. In such models the vertices represent the position sites where the walker can stand, and the edges represents the connection between the vertices, marking sites where the walker could possibly move.
Given a regular finite one-dimensional graph (or lattice) $G(V,E)$, the Hilbert space
$\mathcal{H}$ of a discrete time quantum walk is defined as a composite one
\begin{equation}
\mathcal{H} = \mathcal{H}_P \otimes \mathcal{H}_C\,,
\end{equation}
where the so-called position space $\mathcal{H}_P$
is spanned by state vectors corresponding to vertices $V$ of the graph - thus the positions where the walker can reside. Coin space
 $\mathcal{H}_C$ is spanned by state vectors corresponding to directions labeling nearest neighbor steps (connections), \textit{i. e.} $| L \rangle_C, | R \rangle_C$ for one-dimensional graphs, representing step directions left and right accordingly.
Throughout this paper we use the shorthand $ | x, c \rangle = |x \rangle_P \otimes | c \rangle_C$ for denoting state vectors of the system. With $\mathcal{B(H)}$ we will denote the (Hilbert) space of operators acting on the Hilbert space $\mathcal{H}$. Here, we stress that the graphs we consider are always finite.

Suppose, that due to some errors in the hopping mechanism it is possible that the particle cannot pass through an edge during an unit time interval (discrete step) . We address such an edge to be "broken" during that unit time step. Naturally, in the complementary case the particle is free to pass through the edge. We call such an edge "perfect". We assume that the error causing the edges to be broken or perfect is of statistical nature.  Thus, we assign a probability $p_{\ell}$ to every edge $\ell \in E$ of the graph representing the probability of the edge being perfect during an unit time step. In this way the complementary $1-p_{\ell}$ is the probability of the edge being broken under unit time. We also assume, that all these probabilities are independent. A natural way to describe a system with such imperfections is to introduce dynamical percolation in the underlying graph:
Before every discrete time step we chose an edge configuration  $\mathcal{K} \subseteq E$ randomly (according to the probabilities $p_{\ell}$), which describes the failures in the hopping mechanism. Broken edges are simply missing from the percolation graph, \textit{i. e.} are not in configuration $\mathcal{K} $.

Let us move on to define the time evolution of the system bearing such spatiotemporal noise.
Through the dynamical percolation (randomly changing edge configurations) classical randomness enter the model. We describe the state of the system using density operator formalism. Thus, a unit step of the stochastic time evolution reflects our lack of knowledge about the actual (random) edge configuration $\mathcal{K}$ , \textit{i. e.} an unit step is an incoherent mixture of different coherent time evolutions:
\begin{equation}
\rho(n+1) = \sum_{\mathcal{K}} \pi_{\mathcal{K}} U_{\mathcal{K}} \rho(n) U_{\mathcal{K}}^{\dagger} \equiv \Phi( \rho(n) )\,.
\label{superop:timeevo}
\end{equation}
Here, we defined the linear superoperator $\Phi$ acting on the set of density operators. $\pi_{\mathcal{K}}$ represents
the probability that a given edge configuration $\mathcal{K}$ occurs:
\begin{equation}
\pi_{\mathcal{K}} =  \left\{ \prod_{\ell \in  \mathcal{K}} p_{\ell} \right\} \left\{ \prod_{\ell \not{\in} \mathcal{K}} (1 - p_{\ell}) \right\}\,.
\end{equation}
By $U_\mathcal{K}$ we denote the unitary time evolution operator of the QW on the one-dimensional graph with configuration  $\mathcal{K}$.
The superoperator $\Phi$ belongs to the class of random unitary operations (RUO maps) due to its construction.

\begin{figure}
\begin{center}
\includegraphics[width=0.95\textwidth]{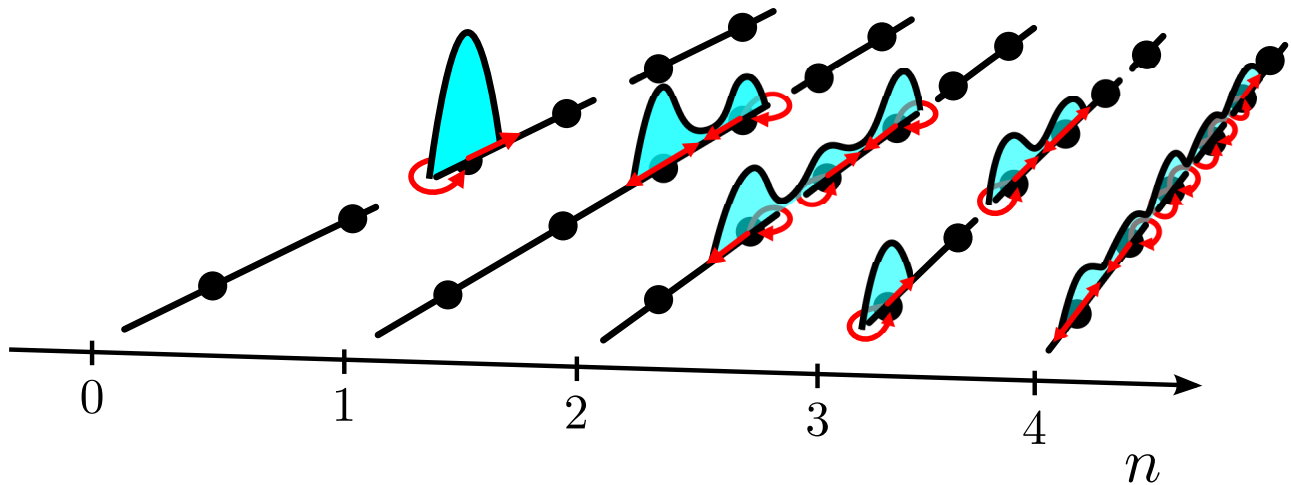}
\end{center}
\caption{
A quantum walk on a one-dimensional graph $G(V,E)$ with dynamical percolation. All the edges $\ell \in E$ of the graph have a corresponding probability $p_{\ell}$ that they might be missing under a discrete time evolution step. We emphasize, that in the figure we show one possible unitary trajectory of the system, \textit{i. e.} the quantum state of the walk is pure. However, due to the (inherently hidden) random nature of the system, the complete, stochastic time evolution consists of the statistical mixture of such unitary trajectories (\ref{superop:timeevo}). The red arrows show the directions where the wave packet can spread (hop) in the next step, while the red loops correspond to the case when the wave packet cannot pass through a missing (broken) edge --- it will suffer a reflection on its internal coin degree of freedom.
}
\label{fig:scheme}  
\end{figure}

The definition of the unitary $U_\mathcal{K}$ is ambiguous since the time evolution of QWs on imperfect (percolation) graphs could be defined in several ways. In the present article we define $U_\mathcal{K}$ by introducing the concept of reflection. Whenever the particle faces a broken (missing) edge, which it cannot pass, it stays at its current place, but suffers a reflection in its internal degree of freedom. We describe this reflection by an off-diagonal unitary matrix $R$. For one-dimensional systems we will use $\sigma_x$ as $R$. We note, that choosing another reflection operators might change the behavior of the system significantly.
Let us now define the unitary evolution for a given configuration $\mathcal{K}$:
\begin{equation}
 U_{\mathcal{K}} = S_\mathcal{K} \cdot (I_P \otimes C)\,,
 \label{unitary_time_evo}
\end{equation}
where
\begin{equation}
S_\mathcal{K}  =   \sum_{(x,x \oplus c)  \in K}  | x \oplus c \rangle_P\langle x |_P \otimes | c \rangle_C\langle c|_C \, +  \sum_{(x,x \oplus c)  \not{\in} K}   | x \rangle_P\langle x |_P \otimes | c \rangle_C\langle c|_C R\,,
\label{PercolationStep}
\end{equation}
is the so-called step operator. The first term  describes the coin dependent shift (hopping) when the edge is perfect: with $x \oplus c$ we denote the nearest neighbor of $x$ in the direction $c$. The second term corresponds to the case when the actual edge is broken: the internal coin degree of freedom suffers reflection by operator $R$ but the particle stays at the same place.
The unitary coin operator corresponds to the coin throw. For one-dimensional lattices we choose $C \in SU(2)$.  Note, that in equation (\ref{unitary_time_evo}) operator $I_P \otimes C$ is spatially homogeneous, thus the configuration dependent $S_{\mathcal{K}}$is the only possible source of inhomogeneousity. A single unitary trajectory of a quantum walk on a dynamical percolation graph is illustrated on figure \ref{fig:scheme}.

Finally, the discrete evolution of the quantum walker on the percolation graph is described by repeated application of the single step of evolution $\Phi$ (See equation (\ref{superop:timeevo}) ), \textit{i. e.} after $n$-th step of evolution the walker initially at the state $\rho_0$ will be found in the state
\begin{equation}
\label{def_iterated_evolution}
\rho(n) = \Phi^n(\rho_0).
\end{equation}
In the next section we review our method of solving the asymptotic dynamics of such evolutions.

%%%%%%%%%%%%%%%%%%%%%%%%%%%%%
\section{Asymptotic dynamics of walks on percolation graphs}
%%%%%%%%%%%%%%%%%%%%%%%%%%%%%
\label{asymptotics}
Both the unitary and the percolative coined quantum walks can be viewed as repeated iterations of one single step. In the case of the percolative quantum walk, there is a random choice of broken edges in each step. For such an open system, each step can be different in a certain realization of the process, nevertheless statistically speaking one can view the process as an iteration of the same step on the density operator of the system. This fact is expressed by the repeated application of the time-independent superoperator introduced in the previous section. The analysis of the dynamics for a percolative quantum walk  is in general more involved than the analysis of the corresponding unperturbed unitary walk. In the latter case, the closed discrete evolution can be described by the iteration of a single unitary operator which commutes with its adjoint. There are two advantages of having a unitary generator at hand. First, it can be diagonalized and, second, we can always choose an orthonormal basis formed by its eigenvectors. In contrast, for the open evolution we have a generator $\Phi$ of equation (\ref{superop:timeevo}), a superoperator acting on density operators. Such a superoperator is not necessarily normal, \textit{i. e.} it does not commute with its adjoint operator, therefore one may not be able to simply diagonalize it in some orthonormal basis.
One can still find the solution for the asymptotic dynamics both for iterated random unitary dynamics \cite{Novotny2009,Novotny2010} and for much more general quantum evolutions \cite{Novotny2012} as well. In the following we outline the key results of this theory.

%%%%%%%%%%%%%%%%%%%%%%%%%%%%%
\subsection{Attractor space}
\label{attractorspace}
One step of random unitary evolution is governed by the linear map $\Phi$ acting on the space  $\mathcal{B} (\mathcal{H})$. Instead of diagonalization one can employ the Jordan canonical form of the map $\Phi$. Eigenvalues from the spectrum $\sigma$ of the map $\Phi$ satisfy $|\lambda| \leq 1$ for any $\lambda \in \sigma$. Jordan blocks which correspond to eigenvalues with modulus one are one-dimensional and the remaining eigenvalues will tend to zero during the iteration, thus, the corresponding Jordan blocks do not play a role asymptotically. Altogether this means that the evolution operator (\ref{def_iterated_evolution}) can be effectively diagonalized in the asymptotic regime. The subspace of the space  $\mathcal{B} (\mathcal{H})$ which does not vanish under many iterations is called the attractor space ($\set{A}$). The attractor space is then spanned by the eigenvectors of the map $\Phi$ associated with eigenvalues of modulus one
\begin{equation}
\set{A} = \bigoplus_{\lambda \in \sigma_1} \set{Ker}\left(\Phi-\lambda I\right)\,.
\label{def_attractor_space}
\end{equation}
Here, we define the attractor spectrum $\sigma_1$ as the asymptotically relevant part in the spectrum  of the RUO map $\Phi$ containing only the eigenvalues which have unit absolute value. One can show that $\set{A}$ is orthogonal to the vanishing part of the space  $\mathcal{B} (\mathcal{H})$.

Random application of unitary operators cannot decrease the von Neumann entropy. In finite systems there is an upper limit for the entropy, consequently, in the asymptotic regime the entropy cannot change.
This property together with the concavity of the entropy implies, that asymptotically any of the randomly chosen unitaries transforms the density operator to the same state, \textit{i. e.} $U_i \rho_{as} U_i^{\dagger} = U_j \rho_{as} U_j^{\dagger}$. In other words, the asymptotic state evolves unitarily, thus the RUO map $\Phi$ is normal on the asymptotic subspace $\set{A}$ and can be diagonalized. The attractors $X \in \set{A}$ corresponding to a given eigenvalue $\lambda$ from the attractor spectrum $\lambda \in \sigma_1$ are determined by the set of equations
\begin{equation}
U_{\mathcal{K}}X = \lambda  XU_{\mathcal{K}}  \quad \forall\,\mathcal{K} \subseteq E\,,\quad |\lambda|=1\,.
\label{attractor_space_matrices}
\end{equation}
One can show that attractors associated with different eigenvalues from the asymptotic spectrum are orthogonal to each other. Consequently, we can choose a basis $X_{\lambda,i}$ of particular kernels of the map $\Phi$ associated with the attractor spectrum. This basis is orthonormal
with respect to the Hilbert-Schmidt scalar product, \textit{i. e.}
\begin{equation}
 \mathrm{Tr} \left( X_{\lambda,i} X_{\lambda',j}^{\dagger} \right) = \delta_{\lambda,\lambda'} \delta_{i,j}\,.
 \label{orthonormalitycondition}
\end{equation}
Indices $i$ and $j$ refer to a possible degeneration of the given eigenvalues $\lambda, \lambda' \in \sigma_1$. Finally, the asymptotic dynamics can be readily determined with the following formula
\begin{equation}
\rho_{as} (n) = \sum_{\lambda \in \sigma_1,i} \lambda^{n} X_{\lambda, i } \cdot \mathrm{Tr} \left( \rho_0 X_{\lambda, i}^{\dagger} \right) \quad \text{when}\quad  n \gg 0\,.
\label{asymptotic_density_operator}
\end{equation}
Here, the phases of eigenvalues $\lambda$ are responsible for the appearance
of non-monotonous asymptotics, \textit{e. g.} limit cycles.

It is important to stress, that equations (\ref{attractor_space_matrices}) do not depend on the probability distribution $\{\pi_{\mathcal{K}}\}$, thus,
the asymptotic behavior of dynamics (\ref{asymptotic_density_operator}) generated by RUO maps is insensitive to the actual $p_{\ell}$ probabilities of errors \cite{Novotny2010}, except in the extremal cases when some $p_{\ell} = 1$ or $0$.
In percolative systems sometimes there exists a critical value of the probability, around which a phase transition occurs in the system. A direct consequence of the insensitivity of the asymptotics dynamics to the particular value of the probability is that the asymptotic dynamics cannot reflect signatures of such a phase transition. 

%%%%%%%%%%%%%%%%%%%%%%%%%%%%%
\subsection{Solution by separation}
\label{sec:separation}
In general, determining the attractor space via equation (\ref{attractor_space_matrices}) is a demanding task. However, it can be simplified considerably with the use of methods we introduced in our previous papers \cite{Kollar2012,Kollar2014}.
We review the essential steps of these methods in the following.

By using the definition in (\ref{unitary_time_evo}) conditions of (\ref{attractor_space_matrices}) take the form
\begin{equation}
S_{\mathcal{K}} \left( I_P \otimes C \right) X \left( I_P \otimes C^{\dagger} \right) S_{\mathcal{K}}^{\dagger} = \lambda X\,,
\end{equation}
which we immediately rewrite into
\begin{equation}
\label{eq:attractor_walk_gen}
\lambda S_{\mathcal{K}}^{\dagger} X S_{\mathcal{K}}  = \left( I_P \otimes C \right) X \left( I_P \otimes C^{\dagger} \right)\,,
\end{equation}
which must be satisfied for all $\mathcal{K} \subseteq E$ with $|\lambda| =1$. A closer look at the latter formula reveals that the right hand side does not depend on edge configurations and the left hand side does not depend on the coin operator. Consequently we can split  solution of equations (\ref{eq:attractor_walk_gen}) into simultaneous solution of ``shift conditions"
\begin{equation}
S_{\mathcal{K}} S_{\mathcal{K'}}^{\dagger} X S_{\mathcal{K'}} S_{\mathcal{K}}^{\dagger} =   X \quad
\forall \, \mathcal{K'}, \mathcal{K} \subseteq E\,
\label{shift_conditions}
\end{equation}
and equation
\begin{equation}
 (I_P \otimes RC)  X  (I_P \otimes C^{\dagger} R^{\dagger}) = \lambda  X,
 \label{coin_condition}
\end{equation}
which corresponds to a single concrete configuration: when all edges of the graph are missing. The reason behind choosing this actual configuration is to make the last coin dependent equation local in position. Using the coin block form of the operator $X = \sum_{s,t} |s\>\<t|\otimes  X^{(s,t)}$ one can realize that the equation (\ref{coin_condition}) is equivalent to the set of the same (local) coin block conditions
\begin{equation}
\label{local_coin_condition}
(RC) X^{(s,t)} (RC)^{\dagger} = \lambda X^{(s,t)}
\end{equation}
for each coin block $X^{(s,t)}$. (We intentionally use the notion ``coin block" because each matrix $X^{(s,t)}$ is defined on the coin Hilbert space $\mathcal H_C$.) Employing the isomorphism $\< \vec x^{(s,t)}|c,d\> \equiv \<c| X^{(s,t)}|d\>$ we can turn (\ref{local_coin_condition}) into an eigenvalue problem of the operator $RC$
\begin{equation}
\label{local_coin_condition_vec}
(RC)\otimes (RC)^{*} \vec x^{(s,t)} = \lambda \vec x^{(s,t)}.
\end{equation}

One can easily infer the meaning of both sets of equations (\ref{shift_conditions}) and (\ref{local_coin_condition}). The shift conditions represent the topology of the system, whereas the coin block condition determines the possible members of the attractor spectrum $\sigma_1$ ( \textit{i. e.} $| \lambda | = 1$ ) and the structure of each corresponding internal coin block. This gives a natural way to solve the problem of searching for attractors. First, one can use the block coin condition (\ref{local_coin_condition}), to determine possible eigenvalues of the attractor spectrum and general form of its corresponding coin block. Next, the whole attractor $X$ have to be built from the particular coin blocks in agreement with shift conditions (\ref{shift_conditions}). In this way, the orthonormal basis $X_{\lambda,i}$ of the attractor space $\set{A}$ can be obtained.

Solving the asymptotics based on separation is a method available for general percolative quantum walks. The disadvantage of this method rests in the fact that attractors are not necessarily density operators, but general elements of the space $\mathcal{B} (\mathcal{H})$. Therefore, the analysis can become quite cumbersome. However, in certain cases, the whole attractor space or its relevant part can be constructed directly using a procedure based on finding pure common eigenstates \cite{Kollar2014}. In the following we review this method, giving its application to the percolative quantum walk.

%%%%%%%%%%%%%%%%%%%%%%%%%%%%%
\subsection{Pure state procedure}
\label{purestatemethod}

In general the attractor space consisting of the $X$ matrices provides an abstract solution for the asymptotic problem, but the $X$ matrices themselves are not necessarily physical, the physically valid asymptotic density operators form only a subset of the attractor space. In order to gain more insight of the possible form of asymptotic density operators we can try to follow a different way to construct them from pure quantum states. Our starting point is the fact that general attractor matrices $X$ of equation (\ref{attractor_space_matrices})  evolve unitarily during the random unitary time evolution $\Phi$ as we discussed in section \ref{attractorspace}. 

Let us consider pure states which are eigenstates of all the possible unitaries $U_{\mathcal{K}}$ in $\Phi$. We refer to these states simply as ``common eigenstates". Obviously, these common eigenstates if written in a density operator form provide a physically valid member of the attractor space. Surprisingly, as we will show it below, the procedure based on finding these states can be very fruitful to construct a substantial part of the attractor space.

Common eigenstates of the unitary operators $U_{\mathcal{K}}$ corresponding to the eigenvalue $\alpha$ can be found as solutions of the set of equations
\begin{equation}
U_{\mathcal{K}} | \psi \rangle = \alpha | \psi \rangle\quad\forall\,\mathcal{K} \subseteq E\,.
\label{commoneigenstates}
\end{equation}
Let us choose an orthonormal basis of common eigenstates $\{ | \phi_{\alpha, i_{\alpha}} \rangle \}$, where index $i$ refers to possible degeneracies. It is apparent that any linear combination
\begin{equation}
Y = \sum_{\alpha \beta^{*} = \lambda, i_{\alpha}, i_{\beta}} A^{\alpha, i_{\alpha}}_{\beta, i_{\beta}}  | \phi_{\alpha,  i_{\alpha}} \rangle \langle \phi_{\beta, i_{\beta}} |\,,
\label{pureattractors}
\end{equation}
with a fixed eigenvalue product $\alpha \beta^{*} = \lambda$ constitute an attractor corresponding to eigenvalue $\lambda$.
Indeed, such attractors by their construction satisfy equations
\begin{equation}
U_{\mathcal{K}} Y = \lambda   Y U_{\mathcal{K'}}\quad \textrm{where}\,|\lambda| = 1\quad\forall \,\mathcal{K, K'}\,.
\label{pureattractor_conditions}
\end{equation}
Remarkably, as was shown in \cite{Kollar2014}, the opposite statement is also true: Any attractor which follows condition (\ref{pureattractor_conditions}) can be written as an incoherent mixture of pure common eigenstates (\ref{pureattractors}). We call these attractors p-attractors.

The condition (\ref{pureattractor_conditions}) in contrast with the condition on general attractor space matrices (\ref{attractor_space_matrices}) is more restrictive, because in this case $Y_{\lambda,i}$ must be invariant under the effect of any pair of unitaries $U_{\mathcal{K}}$ and $U_{\mathcal{K'}}$. Therefore, not all attractors can be constructed from pure common eigenstates in general.  For example
the trivial attractor proportional to identity is not a p-attractor, as it breaks the preceding condition (apart from the case of a purely unitary time evolution). Consequently, the attractor space must always contain the span of all p-attractors and identity, as a minimal subspace. In fact, for certain  RUO based evolutions this minimal subspace covers the whole attractor space. In this case, the asymptotic time evolution simplifies considerably:
\begin{equation}
\rho(n)  =  U_{\mathcal{K}}^n \mathcal{P}
\rho_0 \mathcal{P}
\left(U_{\mathcal{K'}}^{\dagger}\right)^n + \tilde{\mathcal{P}}
\frac{\mathrm{Tr} \left\{ \rho_0
\tilde{\mathcal{P}}\right\}}{\mathrm{Tr} \tilde{\mathcal{P}}} \quad \text{where}\quad  n \gg 0\,.
\label{asymptotic_unitary_evolution}
\end{equation}
Here, $P$ is a projection into the subspace of common eigenstates, and $\tilde{\mathcal{P}}$ is its orthogonal
complement satisfying $\mathcal{P} + \tilde{\mathcal{P}}=I$.

Practically speaking, even if there are some nontrivial non-p-attractors, it is convenient to first construct all p-attractors using the easy-to-obtain pure eigenstates. Then, using the general method  one can construct and add the non-p-attractors to complete the whole attractor space. In the following, we employ the toolset we revised here on one-dimensional QWs, and solve the asymptotic dynamics explicitly.

%%%%%%%%%%%%%%%%%%%%%%%%%%%%%
\section{One-dimensional quantum walks --- complete analysis}
%%%%%%%%%%%%%%%%%%%%%%%%%%%%%
\label{1dqws}

%%%%%%%%%%%%%%%%%%%%%%%%%%%%%
\subsection{Constructing the asymptotics}
\label{sec:constructing_asymptotics}

This part is devoted to combine the capabilities of our methods into an efficient analytical tool for determining the attractor space of quantum walks on certain percolation graphs. We study the line graph (chain) and the circle graph (cycle), both consisting of $N = | V |$ vertices. Thus, the position space $\mathcal{H_P}$ is spanned by state vectors indexed by a single non-negative integer. These graphs represent two physically relevant situations: reflecting and periodic boundary conditions. The coin space $\mathcal{H_C}$  on such $2$-regular graphs are two-dimensional, spanned by states $ | L \rangle_C, | R \rangle_C$, corresponding to stepping left and right respectively. We set the reflection operator $R = \sigma_x$ and choose the coin operators from the $SU(2)$ group.

Let us now present our approach step-by-step.
First, we start our analysis with searching for common eigenstates. According to (\ref{commoneigenstates}) they are defined by equations
\begin{equation}
\label{eq_common_eig_states_1D}
S_{\mathcal{K}} \left(I_P \otimes C\right) |\psi\> = \alpha |\psi\>.
\end{equation}
Repeating similar steps as in section \ref{sec:separation}, we separate (\ref{eq_common_eig_states_1D}) into a local coin condition with one chosen edge configuration
\begin{equation}
S_{\mathcal{K}}(I_P \otimes C) | \psi \rangle = \alpha  | \psi \rangle
\label{purecoincondition}
\end{equation}
and the set of shift conditions
\begin{equation}
S_{\mathcal{K'}}  S_{\mathcal{K}}^{\dagger} | \psi \rangle =  | \psi \rangle
\quad\forall\,\mathcal{K, K'} \subseteq E\,.
\label{pureshiftconditions}
\end{equation}
Expanding an arbitrary pure quantum state as $|\psi\> = \sum_{s} |s\> \otimes |\psi^{(s)}\>_C$ we recast (\ref{purecoincondition}) into the set of local and equivalent eigenvalue equations
\begin{equation}
RC |\psi^{(s)}\rangle_C = \alpha | \psi^{(s)} \rangle_C\,.
\label{spectrum}
\end{equation}
Equations (\ref{spectrum}) determine the possible candidates for eigenvalues $\alpha$ associated with common eigenstates (\ref{commoneigenstates}) and also the general structure of internal coin states $|\psi^{(s)}\>$. These internal coin states are then bound to each other via shift conditions (\ref{pureshiftconditions}). In the natural basis $|\psi \rangle = \sum_{s,c} \psi_{s,c} | s, c \rangle$, the latter shift conditions can be rewritten as
\begin{equation}
 \psi_{s,R} = \psi_{s\ominus1,L} \quad \forall s \in V \, ,
 \label{shiftconditions_vec}
\end{equation}
where the boundary conditions must be taken into account. This procedure provides us the whole subspace of common eigenstates (\ref{commoneigenstates}) and via (\ref{pureattractors}) we can easily construct all p-attractors. At this point is crucial to understand which part of the attractor space $\set{A}$ is formed by p-attractors. All p-attractors are determined by equations (\ref{pureattractor_conditions}). Following the same separation steps as earlier, equations (\ref{pureattractor_conditions}) can be rewritten into the local condition (\ref{local_coin_condition}) for coin blocks and the set of shift conditions for p-attractors which in this case take the form
\begin{equation}
\label{eq:p_attractors_shift_condition}
S_{\mathcal{L}} S_{\mathcal{K}}^{\dagger}  Y S_{\mathcal{K'}} S_{\mathcal{L'}}^{\dagger} =  Y \quad\forall\,
\mathcal{K, K', L, L'} \subseteq E\,.
\end{equation}
In fact, the only difference between general and p-attractors  is in the shift conditions, \textit{cf.} equation (\ref{shift_conditions}) and  equation (\ref{eq:p_attractors_shift_condition}).

Let us investigate this difference in more details. For the system under consideration $S_{\mathcal{K}}$ matrices are always permutation matrices, thus for given configurations they define a one-to-one correspondence between matrix elements. Moreover, we only deal with walks with nearest neighbor steps. These properties imply that a single matrix element determines 3 other matrix elements at most. 
Let us denote a matrix element of p-attractor $Y$ in the natural basis as
\begin{equation}
\langle s_1, c_1 | Y | s_2, c_2 \rangle =  Y^{s_1, c_1}_{s_2, c_2}\,.
\label{1d:matrixelements}
\end{equation}
Using this notation we rewrite shift conditions (\ref{eq:p_attractors_shift_condition}) as
\begin{equation}
Y^{s_1\ominus1,L}_{s_2\ominus1,L}  =  Y^{s_1,R}_{s_2,R} =
Y^{s_1,R}_{s_2\ominus1,L}  =  Y^{s_1\ominus1,L}_{s_2,R} \quad \forall s, s_1, s_2 \in V\,.
\label{1d:allelements_pure}
\end{equation}
We repeat, that the latter equation describes the shift condition requirements on the elements of p-attractors.

Now, we turn to the shift condition (\ref{shift_conditions}) on general attractors to find the subtle, but important differences.
Repeating the same separation, we find that if $s_1 = s_2$, the edges $(s_1\ominus 1,s_1)$ and $(s_2\ominus 1,s_2)$ are the same.
In this case, the state vectors corresponding to these elements see the same configuration (shifts). Thus, for the matrix element $X^{s \ominus 1,L}_{s \ominus 1,L}$ the only requirement is
 \begin{equation}
X^{s \ominus 1,L}_{s \ominus 1,L}  = X^{s,R}_{s,R}.
\end{equation}
One can repeat the above presented steps, to determine all requirements imposed on matrix elements of the general attractor $X$. All these conditions can be written in a concise form
\begin{equation}
X^{s_1\ominus1,L}_{s_2\ominus1,L}  =  X^{s_1,R}_{s_2,R} =
X^{s_1,R}_{s_2\ominus1,L}  =  X^{s_1\ominus1,L}_{s_2,R}\,,
\label{1d:allelements}
\end{equation}
when $s_1 \neq s_2$ is satisfied. When $s_1 = s_2 \equiv s$, less restrictive conditions must hold for the matrix elements of general attractors
\begin{eqnarray}
\label{1d:diagonalelements}
X^{s \ominus 1,L}_{s \ominus 1,L} & = & X^{s,R}_{s,R} \\
 \label{1d:antidiagonalelements}
 X^{s,R}_{s\ominus1,L} & = & X^{s\ominus1,L}_{s,R}\,.
\end{eqnarray}
We note, that these less restrictive conditions are the only differences between the conditions on p-attractors and general attractors. Therefore, first one can construct all p-attractors by pure eigenstates. Then, by allowing (\ref{1d:diagonalelements}) and  (\ref{1d:antidiagonalelements}) the attractor space can be expanded with the remaining non-p-attractors.
The different boundary conditions are handled by the shift conditions in all cases. Thus, if $s_{1(2)} \ominus (\oplus) 1$   belongs to a reflecting boundary (in a case of the line graph), the corresponding equations must be omitted from the set of equations defined above.  

In summary, the complete attractor space can be constructed as follows.
First, using (\ref{spectrum}) one can determine the possible $\alpha$ eigenvalues  and corresponding internal coin states. Second, employing vector shift conditions (\ref{shiftconditions_vec})  all common pure eigenstates can be constructed. Using an orthogonalization process, a corresponding orthonormal basis must be formed from the eigenstates. Next, according to (\ref{pureattractors}) all p-attractors can be constructed, along with the corresponding $\lambda \in \sigma_1$ superoperator eigenvalues. Then, by allowing the general constraints (\ref{1d:diagonalelements}) and (\ref{1d:antidiagonalelements}), the attractor space must be extended to non p-attractors.  In other words, we are looking for the missing attractors, which do not follow shift condition (\ref{1d:allelements_pure}) for the diagonal elements. Instead, the less restrictive conditions (\ref{1d:diagonalelements}) and (\ref{1d:antidiagonalelements}) must be met. In this way at least one additional attractor, proportional to identity, will be found, which is the trivial solution.

%%%%%%%%%%%%%%%%%%%%%%%%%%%%%
\subsection{Explicit solutions}
\label{sec:explicitsolutions}

In \cite{Kollar2012} the structure of attractor space for the percolation line (chain) and cycle was presented for a two-parameter coin class. In this class, the asymptotic position distribution was found to be always uniform. However, the method presented above helps us to go beyond the two-parameter coin class and explicitly solve the complete $SU(2)$ problem. This allows us to find solutions breaking the asymptotic uniformity. Employing the constructive pure eigenstate procedure one can study the asymptotic dynamics in details, and with an additional physical insight. In particular, this approach reveals the existence of edge states, \textit{i. e.} eigenstates which are exponentially localized at the boundaries of the system (in position).

Let us parametrize coins from the $SU(2)$ group as
\begin{equation}
 C(\alpha,\beta,\gamma) = \left(
\begin{array}{cc}
 \left(e^{i (\alpha + \gamma) }-e^{i( \gamma - \alpha) } \right) \cos \beta  \sin \beta & e^{-i \alpha } \cos ^2\beta +e^{i \alpha } \sin ^2\beta \\
 e^{i \alpha } \cos ^2\beta +e^{-i \alpha } \sin ^2\beta  & \left(e^{i (\alpha -\gamma) }-e^{-i (\alpha + \gamma) } \right) \cos \beta  \sin \beta  \\
\end{array}
\right)
\label{def:coin:1D}
\end{equation}
with $\beta \neq k \cdot \pi/2 \,|\, k \in \Intgrs$ and $\alpha \neq k \cdot \pi \,|\, k \in \Intgrs$. Thus we exclude coins leading to a trivial scenario in which the resulting dynamics relabels quantum states merely and does not invoke any interference effect. In fact, such special cases without any quantum interferences represent a purely classical process, thus can be solved using a classical stochastic description, which are not in the scope of the present paper.

To begin, we solve (\ref{spectrum}) to gain the possible eigenvalues of common eigenstates and its associated local coin states. The spectrum of the matrix $\sigma_x C(\alpha,\beta,\gamma)$ is $\{e^{i\alpha},
e^{-i\alpha}\}$ with corresponding eigenvectors $|v_1
\rangle_C=\left(\cos \beta, e^{i \gamma} \sin \beta \right)^{T}$, $|v_2
\rangle_C=\left(\sin \beta, -e^{i \gamma} \cos \beta \right)^{T}$. Equipped with this knowledge, by employing (\ref{shiftconditions_vec}) one can construct the following orthonormal basis of common eigenstates for a percolation chain
\begin{eqnarray}
\label{1D_common_eigenstates_1}
|\phi_1 \rangle &=&\left( \frac{(\cot \beta)^2 -1}{(\cot\beta)^{2N} -1} \right)^{1/2} \left\{ \sum_{s=0}^{N-1} (\cot\beta)^s e^{-i\gamma s} | s\rangle_P \otimes |v_1\rangle_C \right\} \,, \\
\label{1D_common_eigenstates_2}
|\phi_2 \rangle &=&\left( \frac{(\tan \beta)^2 -1}{(\tan\beta)^{2N} -1} \right)^{1/2} \left\{ \sum_{s=0}^{N-1} (-\tan\beta)^s e^{-i\gamma s} | s\rangle_P \otimes |v_2\rangle_C \right\}
\end{eqnarray}
with respect to the  spectrum $\{e^{i\alpha}, e^{-i\alpha}\}$. The proof is the following. Any common eigenstate corresponding to eigenvalue $e^{i\alpha}$ can be be written in the form $|\phi \rangle =\sum_s a_s |s \rangle_P \otimes |v_1\rangle_C$. Using shift conditions (\ref{shiftconditions_vec}) we get $a_{s+1}= e^{-i\gamma}a_s \cot\beta$. Normalization yields the single common eigenstate (\ref{1D_common_eigenstates_1}). Following the same steps one can show that (\ref{1D_common_eigenstates_2}) is the second common eigenstate, corresponding to the eigenvalue $e^{-i\alpha}$.
We note, that these eigenstates are exponentially localized at the boundaries of the system (considering the position distribution), for most of the coin operators. Thus, they can be termed as edge states. We discuss this property in more detail in section \ref{sec:edgestates}.

In view of equation (\ref{pureattractors}) each p-attractor can be constructed directly from common eigenstates (\ref{1D_common_eigenstates_1}) and (\ref{1D_common_eigenstates_2}). Thus the p-attractor subspace corresponding to eigenvalue $\lambda_1 = 1$ is two-dimensional with the orthonormal basis $\{|\phi_1\rangle\langle\phi_1|,|\phi_2\rangle \langle \phi_2|\}$ and p-attractor space corresponding to the eigenvalue $\lambda_2=\exp{(2i\alpha)}$ (resp. eigenvalue $\lambda_3=\exp{(-2i\alpha)}$) is one-dimensional with orthonormal basis  $\{|\phi_1\rangle\langle\phi_2|\}$ (resp. $\{|\phi_1\rangle\langle\phi_2|\}$).  In the special case $\alpha=\pi/2$, both eigenvalues $\lambda_2$ and $\lambda_3$ are equal to $-1$ and associated p-attractor subspace is two-dimensional with orthonormal basis $\{|\phi_1\rangle\langle\phi_2|,|\phi_2\rangle \langle \phi_1|\}$.

We find ourselves at the position to ascertain which attractors remained undiscovered by  the pure state method, \textit{i. e.} attractors which are non-p-attractors. In order to answer this question we have to employ the general separation method. We first determine the dimension of each attractor subspace corresponding to a given eigenvalue. This is straightforwad for the attractors associated with $\lambda=\exp{(2i\alpha)}$ for $\alpha \neq \pi/2$.  According to equations (\ref{local_coin_condition}) and (\ref{local_coin_condition_vec}) the general structure of the corresponding coin blocks is one-dimensional
\begin{equation}
X^{(s,t)} = u^{(s,t)}\left( \begin{array}{cc}
\cos{\beta}\sin{\beta} & -e^{i\gamma} \cos^2{\beta} \\
e^{i\gamma} \sin^2{\beta} & -e^{2i\gamma} \cos{\beta}\sin{\beta}
\end{array}
\right).
\end{equation}
Let us assume that the coin block $X^{(0,0)}$ is determined, \textit{i. e.} parameter $u^{(0,0)}$ is set. Shift conditions for attractors (\ref{1d:allelements}), (\ref{1d:diagonalelements}) and (\ref{1d:antidiagonalelements}) determine all other coin blocks. Consequently, we showed that attractor subspace associated with eigenvalue $\lambda=\exp{(2i\alpha)}$ is one-dimensional and thus equal to the subspace of p-attractors. Similarly, one can easily show the same equivalence among general attractors and p-attractors associated with eigenvalue $\lambda=\exp{(-2i\alpha)}$.

We repeat the same procedure, for the $\lambda=1$  eigenvalue. It turns out, that the
 structure of coin blocks is two-dimensional
\begin{equation}
X^{(s,t)} = \left( \begin{array}{cc}
v^{(s,t)}  & c_1 u^{(s,t)} \\
c_1^* u^{(s,t)}  & c_2 u^{(s,t)}+v^{(s,t)}
\end{array}
\right)
\end{equation}
with $c_1=\exp{(-i\gamma)}\sin{\beta}\cos{\beta}$ and $c_2=-\cos{(2\beta)}$. Each coin block is determined by two parameters $u^{(s,t)}$ and $v^{(s,t)}$. Let us assume that the linear parameters $u^{(0,0)}$ and $v^{(0,0)}$ of the diagonal coin block $X^{(0,0)}$ are determined. The less restrictive shift conditions (\ref{1d:diagonalelements}) and (\ref{1d:antidiagonalelements}) of general attractors determine only the linear parameter $u^{(0,1)}$ leaving the parameter $v^{(0,1)}$ free. Once this parameter is set, all other linear parameters are defined via shift conditions (\ref{1d:allelements}), (\ref{1d:diagonalelements}) and (\ref{1d:antidiagonalelements}). The attractor subspace associated with eigenvalue $\lambda=1$ is thus three-dimensional. In order to get all three elements of this subspace it is sufficient to take p-attractors associated with the eigenvalue $\lambda=1$ and add the identity, \textit{i. e.} the trivial solution.

The last difficulty arises in the degenerate case when coin parameter $\alpha=\pi/2$. By employing  (\ref{local_coin_condition}) and (\ref{local_coin_condition_vec}) we find, that the general structure of the coin blocks corresponding to attractors of $\lambda = -1$ is
\begin{equation}
X^{(s,t)} \equiv  D\left(u^{(s,t)},v^{(s,t)}\right) =\left( \begin{array}{cc}
d_1\left(u^{(s,t)}+v^{(s,t)}\right) & d_2^* v^{(s,t)} \\
d_2 u^{(s,t)}  & -d_1\left(u^{(s,t)}+v^{(s,t)}\right)
\end{array}
\right)
\end{equation}
with $d_1=-1/2\tan{(2\beta)}$ and $d_2=\exp{(i\gamma)}$. Repeating the same steps as above, one can find, that the attractor subspace associated with $\lambda=-1$ is three-dimensional. As the subspace of p-attractors associated with eigenvalue $\lambda=-1$ is only two-dimensional, we miss one attractor. In order to construct such attractor we derive a recurrent formula for coin blocks of p-attractors and attractors. Assume, we know one of the coin blocks $X^{(s,t)}$. Then using shift conditions for p-attractors (\ref{1d:allelements_pure}) one can show that neighboring coin blocks of p-attractors are determined by formulas
\begin{eqnarray}
X^{(s,t+1)}&=&D\left(-\frac{d_2}{d_1}u^{(s,t)}-d_1d_2\left(u^{(s,t)}+v^{(s,t)} \right),d_1d_2\left(u^{(s,t)}+v^{(s,t)}\right)   \right), \\
X^{(s+1,t)}&=&D\left(d_1 d_2^*\left(u^{(s,t)}+v^{(s,t)}\right),-\frac{d_2^* }{d_1}v^{(s,t)}-d_1 d_2^*\left(u^{(s,t)}+v^{(s,t)} \right)\right).
\end{eqnarray}
In the view of these relations we can immediately infer two important conclusions. First, all coin blocks of a p-attractor are determined by a single coin block only. Consequently, the subspace of p-attractors associated with $\lambda=-1$ is two-dimensional indeed. Second, if one coin block is zero, then all other coin blocks and the whole p-attractor is inevitably zero. In other words, p-attractors corresponding to $\lambda=-1$ cannot have a zero coin block. On the contrary, for general attractors shift conditions are less restrictive. Thus, formulas for neighbors of diagonal coin blocks $X^{(s,s)}$ read
\begin{eqnarray}
X^{(s,s+1)}&=&D\left(-\frac{d_2}{d_1}u^{(s,s)}-v^{(s,s+1)},v^{(s,s+1)}   \right), \nonumber\\
X^{(s+1,s)}&=&D\left(u^{(s+1,s)},-\frac{d_2^* }{d_1}v^{(s,t)}- u^{(s+1,s)} \right).
\label{1D_degenerate_attractor_recursive}
\end{eqnarray}
We observe that the linear parameters $v^{(s,s+1)}$ and $u^{(s+1,s)}$ are not determined by parameters of the coin block $X^{(s,s+1)}$. Coin blocks $X^{(s,s+1)}$ and $X^{(s+1,s)}$ are bound to each other by the shift condition (\ref{1d:antidiagonalelements}) and thus only one of these two parameters is free in fact. In order to define an attractor $X$ associated with $\lambda=-1$ we have to specify two parameters of its coin block $X^{(0,0)}$ and one parameter $v^{(s,s+1)}$. All other coin blocks are determined by shift conditions for attractors (\ref{1d:allelements}),(\ref{1d:diagonalelements}) and (\ref{1d:antidiagonalelements}). To construct the last missing independent attractor it is sufficient to choose coin block $X^{(0,0)}$ as zero matrix ($u^{(0,0)}=v^{(0,0)}=0$) and set the linear parameter $v^{(s,s+1)}=1$. All rest coin blocks are defined by shift conditions (\ref{1d:allelements}),(\ref{1d:diagonalelements}) and (\ref{1d:antidiagonalelements}). Apparently, the attractor constructed in this recurrent way can not be a p-attractor.

An elegant analytical form of this recurrently constructed missing attractor for $\lambda=-1$ is not known so far. However, using involved calculations we found that the missing attractor can be written in the closed form
\begin{equation}
X_{\pi/2} = X_{-1} F^{\dagger} E F\,,
\label{1D_degenerate_attractor}
\end{equation}
where
\begin{equation}
X_{-1} = \sum_{s=0}^{N-1} \frac{(-1)^s}{\sqrt{2N}} | s \rangle_P \langle s|_P \otimes \left( \begin{array}{rr} 1 & 0 \\ 0 & -1 \end{array} \right)_C\,.
\label{matrixZ}
\end{equation}
$F$ is the discrete Fourier transformation operator acting on position states
\begin{equation}
F =  \sum_{s,t=0}^{N-1} e^{i 2 \pi s t /N} | s \rangle_P \langle t |_P \otimes I_C\,.
\end{equation}
$E$ is the block diagonal matrix
\begin{equation}
 E = \bigoplus_{g=0}^{N-1} \frac{1}{i \sin \left( \frac{2 g \pi}{N} - \gamma \right)  - \cot 2 \beta} \left( \begin{array}{rr} i \sin \left( \frac{2 g \pi}{N} - \gamma \right) &  e^{-i 2 \pi g / N} \cot 2 \beta \\  e^{i 2 \pi g / N} \cot 2 \beta & i \sin \left( \frac{2 g \pi}{N} - \gamma \right) \end{array} \right)\,.
\end{equation}

Let us summarize the complete list of attractors for different boundaries and numbers of sites. First,
we consider the case with coin parameter $\alpha \neq \pi / 2$.
For the percolation line the attractor space is five-dimensional.
\begin{eqnarray}
\left.\begin{array}{l}
Z_1 = |\phi_1\rangle\langle\phi_1|, \\
Z_2 = |\phi_2\rangle\langle\phi_2|, \\
Z_3 = \frac{1}{\sqrt{2N-2}} \left( I_P \otimes I_C -Z_1 - Z_2 \right)
\end{array}\right\} \quad&\textrm{for}&\quad \lambda_1=1 \nonumber\\
 X=|\phi_1\rangle\langle\phi_2|, \quad&\textrm{for}&\quad \lambda_2=e^{2i\alpha} \nonumber \\
\tilde{X} =|\phi_2\rangle\langle\phi_1|, \quad&\textrm{for}&\quad \lambda_3=e^{-2i\alpha} \,.
\label{all1dattractors}
\end{eqnarray}
On the percolation cycle graph the pure eigenstates of the line graph ( $| \phi_1 \rangle$ and $| \phi_2 \rangle$ ) vanish if they are not translationally invariant. Consequently,  $| \phi_1 \rangle$  is a common eigenstate on cycles if  $(\cot \beta)^N e^{-i \gamma N} = 1$ is satisfied. Likewise, for $| \phi_2 \rangle$ the equation $(- \tan \beta)^N e^{-i \gamma N} = 1$ needs to be true. We note, that only on cycles with even number of edges might both eigenstates be available.
For $(\cot \beta)^N e^{-i \gamma N} = 1$ the attractor space of the cycle consists $\{ Z_1, Z_3 \}$, and when $(- \tan \beta)^N e^{-i \gamma N} = 1$ the attractor space is spanned by $\{ Z_2, Z_3 \}$. When both conditions are met, the attractor space of the QW on the percolation cycle is the same as on the percolation line (\ref{all1dattractors}).
We note, that satisfying the above conditions allowing states $| \phi_1 \rangle$ and $| \phi_2 \rangle$ infers, that they will correspond to a flat distribution in position. That is, they will not be edge states exponentially localized at the boundaries of the graph. This property is clearly understandable as the cycle graph has no dedicated boundaries due to the translation invariance.

For the degenerate case $\alpha = \pi / 2$ the additional attractor $X_{\pi / 2}$ (\ref{1D_degenerate_attractor}) is always present on the percolation line. The shift conditions require translation invariance for the attractors on the percolation cycle. $X_{-1}$ is the only building block in the definition of $X_{\pi /2}$ ( \textit{i. e.} equation (\ref{1D_degenerate_attractor}) ), which might restrict such translational invariance, hence $X_{\pi / 2}$ appears in the attractors space of cycles with an even number of vertices only.

The explicit forms of attractors given above confirms two important properties of studied asymptotic dynamics. First, coin blocks of attractors corresponding to $\lambda \neq 1$ (while $|\lambda| =  1$) are strictly traceless, \textit{i. e.} should one trace out the coin degree of freedom, the remaining position density matrix strictly depend on just the $\lambda = 1$ attractors. Consequently,  if one is interested in the position density operator (\textit{e. g.} want to calculate the position distribution), just attractors corresponding to $\lambda = 1$ , \textit{i. e.} $Z_1, Z_2$ and $Z_3$ are sufficient for the calculation.
Second, all position density operators are stationary in time, 
\begin{equation}
 \rho_P (n) = \Tr_C \rho(n) =  \Tr_C \rho(n+1) =  \rho_P (n +1)\quad \text{where} \quad n \gg 1
\label{stationary_in_time}
\end{equation}
Consequently, limit cycles or other non-stationary asymptotic dynamics might be observable only in the coin degree of freedom. Moreover, it can be shown that the stationarity of the asymptotic position density operator holds true for all similar quantum walks \cite{Kollar2014}.

%%%%%%%%%%%%%%%%%%%%%%%%%%%%%
\subsection{Edge states}
\label{sec:edgestates}

It is well known that the position distribution of a classical walk on a connected undirected graph converge to the uniform distribution. This result does not depend neither on our choice of a graph nor on the initial distribution of the walk. In contrast, we found that the asymptotic position distribution can be nonuniform on percolative quantum walks we study, despite the strong decoherence. Moreover, one can observe the existence of the so-called edge states with exponentially decaying position distributions. Indeed, both common eigenstates (\ref{1D_common_eigenstates_1}) and (\ref{1D_common_eigenstates_2}) manifest this behavior. However, we stress that this interesting effect only arises on the line graph --- on the cycle graph such states can not be observed since they do not fulfill the translation invariance required by the periodical boundary conditions.

In order to study this behavior in more details, let us assume the initial state of the walker to be $\rho_0$. After sufficiently many iterations we reach the asymptotic evolution, which according to (\ref{asymptotic_density_operator}) can be written as
\begin{equation}
\rho(n) = O_1 Z_1 + O_2 Z_2 + \frac{1-O_1-O_2}{\sqrt{2N-2}}Z_3 + \mathcal{R}\quad \text{where} \quad n \gg 1\,.
\end{equation}
Here, $O_i$ are the overlap of the initial state $\rho_0$ with common eigenstate (\ref{1D_common_eigenstates_1}) and (\ref{1D_common_eigenstates_2}), 
\textit{i. e.} $O_i = \Tr\{Z_i \rho_0 \}$. They satisfy relation $O_1 + O_2 \leq 1$. The traceless operator $\mathcal{R}$ refers to the overlap of the initial state $\rho_0$ with attractors $X$, $\tilde{X}$ and $X_{\pi / 2}$. However, this part does not contribute to the asymptotic position distribution of the walker, which reads
\begin{equation}
\label{position_distribution}
P(s) = \<s, \Tr_P \{\rho(n) \} s\> = \mathcal{N} \left( O_1 q^{N-1-s} + O_2 q^s\right) + \frac{1-O_1-O_2}{2N}\quad \text{where} \quad n \gg 1\,.
\end{equation}
We defined  $\mathcal{N} = (q-1)/(q^{N}-1 )$ and $q = \tan(\beta)$. The first term in (\ref{position_distribution}) clearly displays the exponentially decreasing behavior from the left edge to the right, and then from a certain minimum an exponentially increasing probability towards to the right edge. The minimum depends on initial overlaps $O_i$. If one of these overlaps is zero one can observe a monotonous exponential behavior.
The second term of the position distribution (\ref{position_distribution}) is constant and might dampen the exponential behavior slightly.

A natural question arises, namely for what coins the exponential localization effect is the most significant. One can easily infer that this occur if $\beta$ is close to values $k \cdot \pi/2 | k \in \Intgrs$. However, these are coins which simply permute the internal coin degrees and do not invoke any interference effects. On the other hand they constitute extreme cases of biased coins for which one could reckon this behavior. Surprisingly, the exponential localization behavior of the asymptotic position distribution is also available for unbiased coins, \textit{i. e.} whose matrix elements has the same amplitude. One can check that these are coins $\mathcal{C}(\alpha,\beta,\gamma)$ which follow condition $\left|\sin(\alpha)\sin(2\beta)\right|=1/\sqrt{2}$. From this condition one can see that the most significant behavior is obtained for $\beta = \pi/8$ and $\alpha = \pi/2$, which corresponds to the coin
\begin{equation}
\mathcal{C}(\alpha=\pi/2,\beta=\pi/8,\gamma) = \frac{i}{\sqrt{2}} \left(
\begin{array}{rr}
e^{i\gamma} & -1  \\
1 & e^{-i\gamma}
\end{array}
\right)\,.
\label{most_exponential_uniform_coin}
\end{equation}
Thus, even for unbiased coins we can get strong exponential behavior of the position distribution (\ref{position_distribution}) with $q=(\tan(\beta))^2=3-2\sqrt{2} \approx 0.1716$.

In the following, we show a step-by-step walkthrough of the methods given in the current section, through an explicit example.

%%%%%%%%%%%%%%%%%%%%%%%%%%%%%
\section{Explicit case study of a one-dimensional QW}
%%%%%%%%%%%%%%%%%%%%%%%%%%%%%
\label{sec:1dexample}

Let us consider the one-dimensional coin operator of equation (\ref{def:coin:1D}) with parameters $\alpha = \frac{Pi}{2},\, \beta=\tan^{-1}\left(\frac{1}{2}\right),\, \gamma = 0$. This gives the following coin matrix:
\begin{equation}
 \mathcal{C} = \frac{i}{5} \left(
\begin{array}{rr}
4 & -3  \\
3 & 4 
\end{array}
\right)\,.
\label{example:coin}
\end{equation}
For the sake of simplicity we will consider a line graph with 4 vertices ($N = 4$). We repeat, that we set the reflection operator $R = \sigma_x$.

%%%%%%%%%%%%%%%%%%%%%%%%%%%%%
\subsection{Attractor space}

To begin, we construct all p-attractors (\ref{pureattractors}) first, which involves finding all pure common eigenstates (\ref{eq_common_eig_states_1D}). We first focus on the case when all edges of the graph are missing: $\mathcal{K} = \{ \}$. In this case the walker is not able to step, only the internal coin degree of freedom is rotated and reflected.
Since this is a possible scenario indeed (all edges can be missing with a finite probability), the asymptotic state (now assumed to be a pure common eigenstate) must be invariant under such time evolution. This results the local eigensystem equation (\ref{spectrum}), which we now repeat in the matrix form:
\begin{equation}
 R \mathcal{C} \vec{v}_{\alpha} = 
 \frac{i}{5} \left(
\begin{array}{rr}
3 & 4  \\
4 & -3 
\end{array}
\right)
 = \alpha \vec{v}_{\alpha}\,.
\end{equation}

\begin{figure}
\begin{center}
\includegraphics[width=0.95\textwidth]{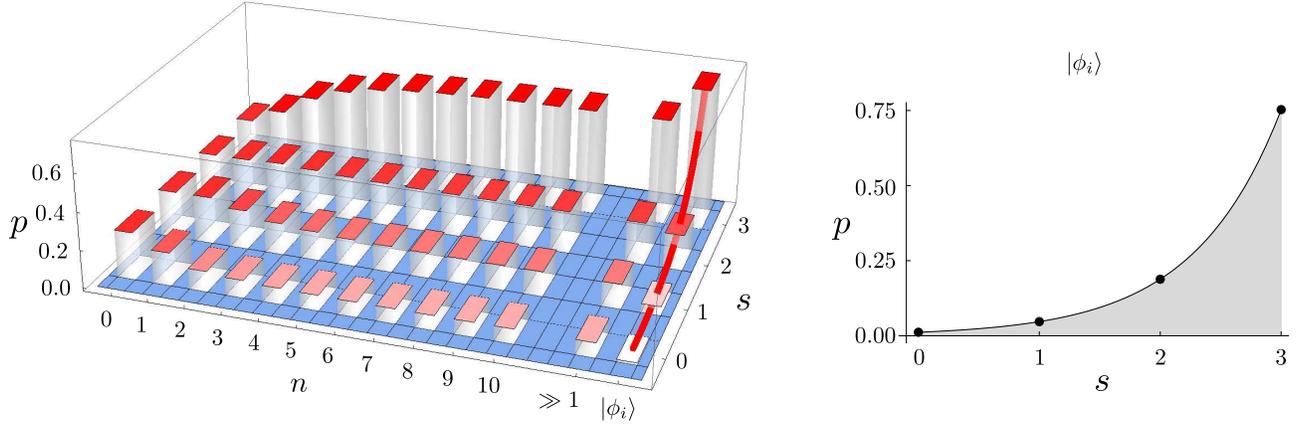}
\end{center}
\caption{
Position probability distribution of the first $10$ steps of the quantum walk examined in section \ref{sec:1dexample}. on the left plot. The quantum walk is performed on a percolation line graph (chain) with 4 vertices, using the coin $\mathcal{C}$ ( see equation (\ref{example:coin})  ). The probability of all edges to be perfect is $p_l = \frac{1}{2}$. The plot corresponds to the walk starting from  $| \psi_0 \rangle$ as in (\ref{psizeros}). These state has a flat distribution in position, however due to the considerable overlap with the edge states $| \phi_i \rangle$  (see equation (\ref{example:phiplusi}) ) the distribution will evolve toward a non-uniform distribution. We show the position distribution of the edge state $ | \phi_i \rangle$ at the corresponding mark on the left plot, and by itself on the right plot respectively. The plot markers of the edge state are connected to emphasize the exponential localization.
The analytically calculated long time (asymptotic) position distribution is shown at the mark $\gg 1$. 
A walk started from the state $| \psi_0' \rangle$  would result a mirrored distribution in position.
We note, that oscillations due to the attractors corresponding to the $\lambda=-1$ superoperator eigenvalue cannot be seen in the position distribution --- this is due to the fact that all asymptotic position distributions are stationary in time (\ref{stationary_in_time}).
}
\label{fig:exampleevo}  
\end{figure}

The solutions are
\begin{eqnarray}
 \vec{v}_{i} & = & \frac{1}{\sqrt{5}}(2,1)^T \quad\text{with eigenvalue}\quad\alpha=i \nonumber\\
 \vec{v}_{-i} & = & \frac{1}{\sqrt{5}}(1,-2)^T \quad\text{with eigenvalue}\quad\alpha=-i\,.
\label{example:coineigenstates}
\end{eqnarray}
This implies, that the form of pure eigenstates corresponding to $\alpha=i$ must be
\begin{equation}
 | \phi_i \rangle = \sum_{j=1}^{4} a_j | j \rangle_P \otimes \left( 2 | L  \rangle_C + | R \rangle_C \right)\,,
\end{equation}
that is, in a vector form
\begin{equation}
 \vec{\phi}_i = \left( 2 a_1, a_1, 2 a_2, a_2, 2 a_3, a_3, 2 a_4, a_4 \right)^T\,.
\end{equation}
Now, we employ the shift conditions  (\ref{shiftconditions_vec}) to eliminate most of the free parameters:
\begin{equation}
 \vec{\phi}_i = \left( 2 a_1, a_1, 4 a_1, 2 a_1, 8 a_1, 4 a_1, 16 a_1, 8 a_1 \right)^T\,.
\end{equation}
Since $ \vec{\phi}_i$ is the vector expansion of a pure quantum state $| \phi_i \rangle$, its norm should be one, and the global phase can be chosen freely. Thus, its final form is
\begin{equation}
 \vec{\phi}_i= \frac{1}{5\sqrt{17}} \left( 2, 1, 4, 2, 8, 4, 16, 8 \right)^T\,.
\label{example:phiplusi}
\end{equation}
Likewise, the other pure common eigenstate corresponding to $\alpha = -i$ can be constructed as:
\begin{equation}
 \vec{\phi}_{-i} = \frac{1}{5\sqrt{17}} \left( 8, -16, -4, 8, 2, -4, -1, 2 \right)^T\,.
\label{example:phiminusi}
\end{equation}
Note, that  $ | \phi_{-i} \rangle$ and $| \phi_{i} \rangle$ are edges states: they are living exponentially localized at the boundaries of the line graph. This property is illustrated on  figure \ref{fig:exampleevo}. Above, we have explicitly derived the common eigenstates of the time evolution. Alternatively, for one-dimensional QWs one could use the formulae (\ref{1D_common_eigenstates_1}) and (\ref{1D_common_eigenstates_2}) to obtain the same result.

The next step is to construct the p-attractors. Using equation (\ref{pureattractors}), the following attractors can be obtained.
\begin{eqnarray}
  \left. \begin{array}{rrr} Z_{1} & = & | \phi_i \rangle \langle \phi_i | \\  Z_{2} & = & | \phi_{-i} \rangle \langle \phi_{-i} |  \end{array} \right\}
  \quad\text{with superoperator eigenvalue}\quad\lambda_1 & = & 1\,, \nonumber\\
  \left. \begin{array}{rrr} X_{1} & = & | \phi_i \rangle \langle \phi_{-i} | \\  X_{2} & = & | \phi_{-i} \rangle \langle \phi_{i} |  \end{array} \right\}
  \quad\text{with superoperator eigenvalue}\quad\lambda_{2} & = & -1\,.
  \label{example:pattractors}
\end{eqnarray}
It is straightforward to see that all these p-attractors are indeed fixed points of the dynamics.
As an example, we show this for $Z_1$. Due to its construction $| \phi_i \rangle$ satisfy:
\begin{equation}
 U_{\mathcal{K}} | \phi_i \rangle = i | \phi_i \rangle \quad \forall\,\mathcal{K} \subseteq E\,.
\end{equation}
Hence, the following equation is always satisfied
\begin{equation}
U_{\mathcal{K}} Z_1   U_{\mathcal{K}}^{\dagger} = U_{\mathcal{K}} | \phi_i \rangle \langle \phi_i |  U_{\mathcal{K}}^{\dagger} = i | \phi_i \rangle \langle \phi_i | (-i) =
| \phi_i \rangle \langle \phi_i | = Z_1\quad\forall\,\mathcal{K} \subseteq E\,,
\end{equation}
which is the definition of an attractor space matrix (\textit{cf.} equation (\ref{attractor_space_matrices}) ), thus, $Z_1$ is indeed a valid attractor. Similarly, it is simple to show that all p-attractors (\ref{example:pattractors}) are elements of the attractors space. Moreover, they automatically satisfy the orthonormality described by equation (\ref{orthonormalitycondition}).

Now, we move on to make our solution complete, \textit{i. e.} find all non-p-attractors. Since the non-p-attractors are usually do not correspond to quantum states, the process is less intuitive, albeit similar to the construction of p-attractors.
To begin, we solve the coin block condition (\ref{local_coin_condition}), \textit{i. e.}:
\begin{equation}
R\mathcal{C} B \mathcal{C}^{\dagger} R^{\dagger} = \lambda B\,.
\label{example:coinblockcondition}
\end{equation}
From the p-attractor analysis, we know, that the set of $|\lambda| = 1$ eigenvalues ( $\sigma_1$ ) contains at least $\pm 1$. As we explicitly solve equation (\ref{example:coinblockcondition}) it turns out, that, there are not any other $\sigma_1$ eigenvalues.
For $\lambda = 1$ the coin block satisfying  equation (\ref{example:coinblockcondition}) has the following form
\begin{equation}
B = \left(
\begin{array}{cc}
a + \frac{3}{2} b & b \\
b & a
\end{array}
\right)\,,
\end{equation}
where $a$ and $b$ are free complex parameters.
Thus, the complete attractor space matrix must have the form
\begin{equation}
Z =  \left(
\begin{array}{cccccccc}
a_1 + \frac{3}{2} b_1 & b_1 & a_2 + \frac{3}{2} b_2 & b_2 & a_3 + \frac{3}{2} b_3 & b_3 & a_4 + \frac{3}{2} b_4 & b_4  \\
b_1 & a_1 & b_2 & a_2 & b_3 & a_3 & b_4 & a_4 \\
a_{5} + \frac{3}{2} b_{5} & b_{5} & a_{6} + \frac{3}{2} b_{6} & b_{6} & a_{7} + \frac{3}{2} b_{7} & b_{7} & a_{8} + \frac{3}{2} b_{8} & b_{8}  \\
b_{5} & a_{5} & b_{6} & a_{6} & b_{7} & a_{7} & b_{8} & a_{8} \\
a_{9} + \frac{3}{2} b_{9} & b_{9} & a_{10} + \frac{3}{2} b_{10} & b_{10} & a_{11} + \frac{3}{2} b_{11} & b_{11} & a_{12} + \frac{3}{2} b_{12} & b_{12}  \\
b_{9} & a_{9} & b_{10} & a_{10} & b_{11} & a_{11} & b_{12} & a_{12} \\
a_{13} + \frac{3}{2} b_{13} & b_{13} & a_{14} + \frac{3}{2} b_{14} & b_{14} & a_{15} + \frac{3}{2} b_{15} & b_{15} & a_{16} + \frac{3}{2} b_{16} & b_{16}  \\
b_{13} & a_{13} & b_{14} & a_{14} & b_{15} & a_{15} & b_{16} & a_{16} \\
\end{array}
\right)\,.
\label{example:X}
\end{equation}
We now compare the shift condition on p-attractors (\ref{1d:allelements_pure}) with the conditions on all attractors (\ref{1d:diagonalelements}), (\ref{1d:antidiagonalelements}) and (\ref{1d:allelements}). The only difference arises from  (\ref{1d:diagonalelements}), (\ref{1d:antidiagonalelements}): the diagonal elements of the general attractors can be chosen independently from the rest of the matrix, \textit{i. e.} we can take any p-attractors and change its diagonal according to (\ref{1d:diagonalelements}) to get a non-p-attractor.
Now, we take a look at $B$: if the off-diagonal elements are zero, the diagonal can be choosen to be nonzero. By joining this two properties an important consequence can be drawn: all non-p-attractors corresponding to coin block $B$ (superoperator eigenvalue $\lambda = 1$) can be chosen as diagonal matrices. Restricting $Z$ of equation (\ref{example:X}) to diagonal matrices and applying (\ref{1d:diagonalelements}) results:
\begin{equation}
Z = a_1 \cdot I_P \otimes I_C \,.
\end{equation}
Which we set in the form
\begin{equation}
Z_3' = \frac{1}{2\sqrt{2}} I
\end{equation}
applying the normalization condition (\ref{orthonormalitycondition}). We note, that this attractor is the trivial one, which is always present in the attractor space: RUO maps are unital.
With proper orthogonalization it takes the form of
\begin{equation}
Z_3 =  \frac{1}{\sqrt{6}}\left( I - | \phi_i \rangle \langle \phi_i | - | \phi_{-i} \rangle \langle \phi_{-i} | \right) =  \frac{1}{\sqrt{6}}\left( I - X_1 - X_2 \right)\,.
\label{X3attractor}
\end{equation}

In a similar manner one could construct all non-p-attractors for the $\lambda=-1$ superoperator eigenvalue. In this exact case, there is a single additional non-p-attractor for $\lambda=-1$: $X_{\pi/2}$ as given in (\ref{1D_degenerate_attractor_recursive}) or (\ref{1D_degenerate_attractor}). This attractor after a proper orthornomalization takes the form of:
\begin{equation}
X_{\pi/2} =  \frac{1}{425 \sqrt{6}} \left(
\begin{array}{rrrrrrrr}
-168 &  126 & -126 & 257 & 68 & -126 & -24 & 68 \\
126 &  168 & -68 & 126 & 24 & -68 & -32  & 24 \\
-126 &  -68 & 168 & -126 & 126 & -257 & -68 & 126 \\
257 &  126 & -126 & -168 & 68 & -126 & -24 & 68 \\
68 &  24 & 126 & 68 & -168 & 126 & -126 & 257 \\
-126 &  -68 & -257 &-126 &126 &168 &-68 & 126 \\
-24 &  -32 & -68 & -24 & -126 & -68  & 168 & -126 \\
68 &  24 & 126 & 68 & 257 & 126 & -126 & -168  
\end{array}
\right)\,.
\end{equation}

%%%%%%%%%%%%%%%%%%%%%%%%%%%%%
\subsection{Employing the asymptotics --- addressing the highest non-uniformity}

We use the analytical results derived in the previous, to explore some interesting properties of the walk under consideration. One might ask the question: Is it possible to achieve a non-flat asymptotic position distribution from a flat initial state, and if the answer is positive, which is the initial state resulting the most non-flat position distribution?

To address this issue, we recall, that in the asymptotic position distribution just the attractors corresponding to the $\lambda = 1$ eigenvalue can play a role. ( This result is discussed in section \ref{sec:explicitsolutions}. ) Hence, we consider $Z_1, Z_2$ and $Z_3$ from equations (\ref{example:pattractors}) and (\ref{X3attractor}). We use the Manhattan distance
$M(p,q) = \sum_i |p_i - q_i|$ to measure the distance of the asymptotic position distribution from the uniform distribution. 

We employ (\ref{asymptotic_density_operator}) to determine the asymptotic position density operator.
\begin{equation}
\rho_P (n) = \Tr_C \left[ Z_1 \cdot \mathrm{Tr} \left( \rho_0 Z_1^{\dagger} \right) + Z_2 \cdot \mathrm{Tr} \left( \rho_0 Z_2^{\dagger} \right) + Z_3 \cdot \mathrm{Tr} \left( \rho_0 Z_3^{\dagger} \right) \right] \quad \text{where} \quad n \gg 1\,.
\end{equation}
Using the linearity of the trace and that $Z_i = Z_i^{\dagger}$ we find
\begin{equation}
\rho_P (n) = \Tr_C \left( Z_1  \right) \cdot \mathrm{Tr} \left( \rho_0 Z_1 \right)  +  \Tr_C \left(  Z_2  \right) \cdot \mathrm{Tr} \left( \rho_0 Z_2 \right) +  \Tr_C \left(  Z_3 \right) \cdot \mathrm{Tr} \left( \rho_0 Z_3 \right) \quad \text{where} \quad n \gg 1 \,.
\end{equation}
We will use the shorthand $O_i$ to denote the overlap of the attractors with the initial state
\begin{equation}
 O_i = Tr \left( \rho_0 Z_i \right)\,.
\end{equation}
We note, that up to a real normalization factor $Z_i$ attractors are all proper density operators, and overlaps $O_i$ are positive real numbers satisfying $\sum_i O_i  \leq 1$, \textit{i. e.} the asymptotic position distribution is a linear combination of the position distributions $\Tr_C \left(  Z_i\right)$. Moreover by looking at the structure of $Z_3$ one might easily calculate its overlap with the initial state as
\begin{equation}
 O_3 =  \frac{1}{\sqrt{6}} \left(1-  O_1 - O_2 \right)\,.
\end{equation}
After the explicit calculation of all $\Tr_C \left(  Z_i\right)$, we can write the Manhattan distance from the uniform distribution in the concise form
\begin{eqnarray}
M  & = & | O_1 \frac{1}{85} + O_2 \frac{64}{85} + O_3 \frac{7\sqrt{3}}{34 \sqrt{2}} - \frac{1}{4}| +  | O_1 \frac{4}{85} + O_2 \frac{16}{85}  + O_3 \frac{5\sqrt{3}}{17 \sqrt{2}} - \frac{1}{4}|  +
\nonumber\\  && | O_1 \frac{16}{85} + O_2 \frac{4}{85} + O_3 \frac{5\sqrt{3}}{17 \sqrt{2}} - \frac{1}{4}| +  | O_1 \frac{64}{85} + O_2 \frac{1}{85} + O_3 \frac{7\sqrt{3}}{34 \sqrt{2}} - \frac{1}{4}|\,.
\end{eqnarray}
One have to maximize the last expression over the set of initial states $\rho_0$ corresponding to a flat position distribution. Since the number of parameters is rather high in such states, it is favorable to perform such task numerically. We found states
\begin{eqnarray}
 | \psi_0 \rangle & = & \frac{1}{\sqrt{20}} \sum_{s=0}^3 | s \rangle_P \otimes \left( 2 | L \rangle_C + | R \rangle_C \right)\nonumber\\
 | \psi_0' \rangle & = & \frac{1}{\sqrt{20}} \sum_{s=0}^3 (-1)^s | s \rangle_P \otimes \left(  | L \rangle_C - 2 | R \rangle_C \right)
\label{psizeros}
\end{eqnarray}
which has a maximal distance of $M = \frac{735}{1156} \approx 0.636$ asymptotically. The time evolution and asymptotics of these states are illustrated in figure \ref{fig:exampleevo}.

%%%%%%%%%%%%%%%%%%%%%%%%%%%%%
\section{Two-dimensional quantum walks}
%%%%%%%%%%%%%%%%%%%%%%%%%%%%%
\label{Sec:2DQWS}

Two-dimensional quantum walks \cite{Mackay2002,Tregenna2003} are straightforward generalizations of one-dimensional quantum walks, extending the basic definitions to two-dimensional graph. However, due to the much broader selection of coins available, \textit{i. e.} $SU(4)$ for 4-regular two-dimensional lattices, the possible effects show a much richer variety. In fact, the complete $SU(4)$ family of coins and the corresponding behaviors are yet to be discovered. Similarly, the percolation problem on two-dimensional graphs are much more complicated, but on the same time bear new phenomena, \textit{e. g.} phase transitions at a critical threshold.
One can except that joining these two fields might exhibit interesting novel results.

In our earlier work \cite{Kollar2014} of QWs on two-dimensional dynamical percolation graphs we indeed discovered some new phenomena: asymptotic position inhomogenity, breaking of the directional symmetry and the survival of the localization (trapping) effect. In the following we give a brief review of the methods, and we highlight some interesting results.

%%%%%%%%%%%%%%%%%%%%%%%%%%%%%
\subsection{Description and asymptotics}

To begin, we note that the general observations on percolative walks (See section \ref{asymptotics}) hold true, and we will refer to them.
The Hilbert space of the two-dimensional QWs is a composite one: $\mathcal{H} = \mathcal{H}_P \otimes \mathcal{H}_C$, where the position space $\mathcal{H}_P$ is spanned by states corresponding to the vertices of a two-dimensional Cartesian lattice with $M \otimes N$ sites, and the coin space $\mathcal{H}_C$ is spanned by vectors corresponding to nearest neighbor steps: $ | L \rangle, | D \rangle, | U \rangle, | R \rangle$ --- we expand all 4-by-4 matrices on this basis respectively.
A single step of the time evolution on a percolation graph is given by equations (\ref{unitary_time_evo}) and (\ref{PercolationStep}). We define the reflection operator as $R = \sigma_x \otimes \sigma_x$.

To solve the asymptotic dynamics of such a system, first one have to find all p-attractors --- in analogy with the one-dimensional case. This can be done by employing equation (\ref{commoneigenstates}) as
\begin{equation}
\label{eq_common_eig_states_2D}
S_{\mathcal{K}} \left(I_P \otimes C\right) |\psi\> = \alpha |\psi\>.
\end{equation}
Which we again separate into a local coin condition with one chosen edge configuration
\begin{equation}
S_{\mathcal{K}}(I_P \otimes C) | \psi \rangle = \alpha  | \psi \rangle
\label{purecoincondition2D}
\end{equation}
and to a set of shift conditions
\begin{equation}
S_{\mathcal{K'}}  S_{\mathcal{K}}^{\dagger} | \psi \rangle =  | \psi \rangle
\quad\forall\,\mathcal{K, K'} \subseteq E\,.
\label{pureshiftconditions2D}
\end{equation}
Let us expand a pure states on the natural basis $|\psi \rangle = \sum_{s,t,c} \psi_{s,t,c} | s, t \rangle_P \otimes |  c \rangle$. Employing this notation, we can rewrite the shift conditions (\ref{pureshiftconditions2D}) as
\begin{equation}
\left. \begin{array}{rrr}
 \psi_{s,t,R} &=& \psi_{s\ominus1,t,L}  \\
 \psi_{s,t,U} &=& \psi_{s,t\ominus1,D}  
 \end{array} \right\}
 \quad \forall (s,t) \in V \, .
 \label{shiftconditions_vec2D}
\end{equation}
Here we note, that in the shift conditions the boundary conditions must be taken into account.
In \cite{Kollar2014} we analyzed periodic (torus) and reflective (carpet) boundary conditions, however the method can naturally be extended to other topologies as well. 

After one successfully constructed all pure common eigenstates, by employing (\ref{pureattractors}) all p-attractors can be build in a straightforward way. 
As in the one-dimensional case, all such p-attractors satisfy the shift condition
\begin{equation}
\label{eq:p_attractors_shift_condition2D}
S_{\mathcal{L}} S_{\mathcal{K}}^{\dagger}  Y S_{\mathcal{K'}} S_{\mathcal{L'}}^{\dagger} =  Y \quad\forall\,
\mathcal{K, K', L, L'} \subseteq E\,.
\end{equation}
However, one can see that general attractors must satisfy a less strict condition
\begin{equation}
\label{eq:general_attractors_shift_condition2D}
S_{\mathcal{K'}} S_{\mathcal{K}}^{\dagger}  Y S_{\mathcal{K}} S_{\mathcal{K'}}^{\dagger} =  Y \quad\forall\,
\mathcal{K, K'} \subseteq E\,.
\end{equation}
Consequently, one can investigate the differences between the two latter set of shift conditions, and construct all missing non-p-attractors. The whole process is analogous to the method we give in section \ref{1dqws}. However, since the dimension and the possible degeneracies in the system are higher, the analysis might be much more involving. The explicit procedure is shown in detail in \cite{Kollar2014}. 
In the next section we review a particularly interesting special case:  the walk driven by the Grover diffusion operator.

%%%%%%%%%%%%%%%%%%%%%%%%%%%%%
\subsection{Example: The Grover-walk}
\label{sec:grover}

Two-dimensional quantum walks driven by the Grover coin
\begin{equation}
 G = \frac{1}{2} \left(
 \begin{array}{rrrr}
 -1 &  1 &  1 &  1 \\
  1 & -1 &  1 &  1 \\
  1 &  1 & -1 &  1 \\
  1 &  1 &  1 & -1 
 \end{array} 
 \right)
\end{equation}
gained a considerable interest in the literature, due to their use in quantum search \cite{Shenvi2003} and also by exhibiting the property of trapping (localization) \cite{Inui2004}. This latter phenomena is the inability of some part of the wave function to leave its initial position due to destructive interference of the outgoing waves \cite{Stefanak2010}. That is, a walker started from a localized initial state always can be found at its initial position with a finite probability. 

\begin{figure}
\begin{center}
\includegraphics[width=0.99\textwidth]{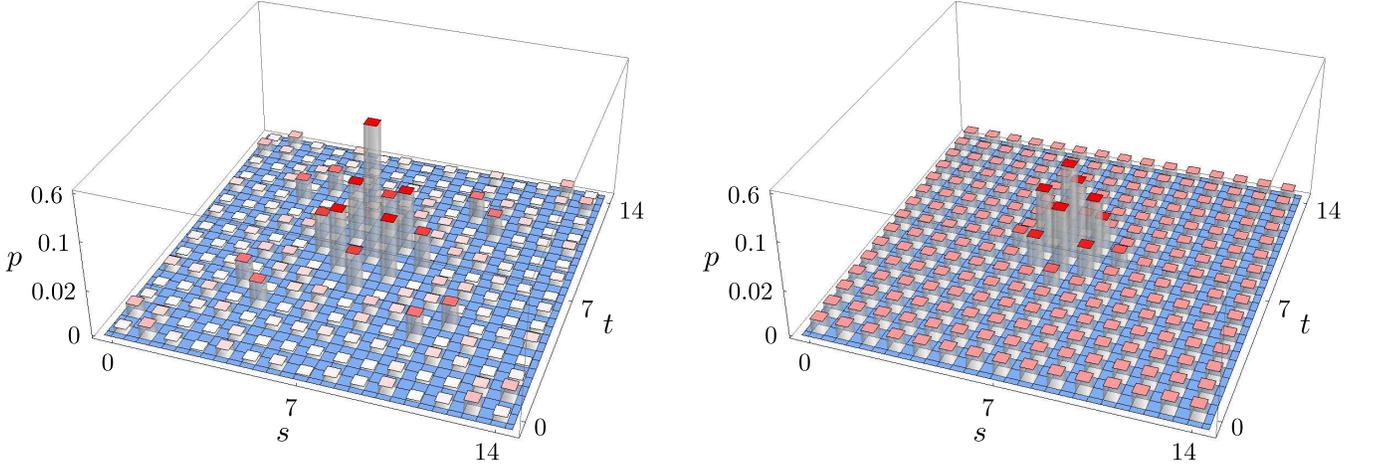}
\end{center}
\caption{
Trapping (localization) of Grover walks (see section \ref{sec:grover})  on the $15 \times 15$ torus (periodic boundaries). The left plot
shows the distribution after 1000 steps if unitary time evolution on a perfect graph.
The right plot shows an asymptotic position distribution of the Grover walk on a percolation graph.
Both walks are started from the initial state $ | 7,7 \rangle_P \otimes \frac{1}{2}( |L \rangle + |
D \rangle + | U \rangle + | R \rangle )$. We used logarithmic scale on the vertical axe, to make the structure of the distributions visible.
The localization property is observed both for the closed and open system dynamics. This effect is due to the common eigenstates of the system with finite support (\textit{cf.} equation (\ref{locgroeigs}) ).
We note, that in the percolation (right plot) case the peak is not as high as for the unperturbed walk.
}
\label{fig:grover}  
\end{figure}

In the following we show the attractor space of the Grover
walk. The common eigenstates defined via equation (\ref{eq_common_eig_states_2D}) have the explicit form of
\begin{eqnarray}
\label{locgroeigs}
 | \phi_1 \rangle & = &  \frac{1}{\sqrt{4MN}} \sum_{s=0}^{M-1} \sum_{t=0}^{N-1} | s, t \rangle_P \otimes | v_1 \rangle_C\,, \\
  | \phi_2 (s,t) \rangle & = & \frac{1}{\sqrt{8}} \big\{ | s, t \rangle_P \otimes | v_2 \rangle_C +  | s, t \oplus 1 \rangle_P \otimes \left( | v_2 \rangle_C + | v_3 \rangle_C \right) \nonumber\\ &&
  + | s \oplus 1, t \rangle_P \otimes \left( | v_2 \rangle_C + | v_4 \rangle_C \right) \nonumber\\ &&  +  | s \oplus 1, t \oplus 1 \rangle_P \otimes \left(  | v_2 \rangle_C +  | v_3 \rangle_C  +   | v_4 \rangle_C   \right) \big\} \,, \label{grover_loc_eigenstates} \\
 | \phi_3 (s) \rangle & = & \sum_{t=0}^{N-1} \frac{(-1)^t}{\sqrt{2N}} | s, t \rangle_P \otimes | v_3 \rangle_C\,, \label{grover_phi3} \\
 | \phi_4 (t) \rangle & = & \sum_{s=0}^{M-1} \frac{(-1)^s}{\sqrt{2N}} | s, t \rangle_P \otimes | v_4 \rangle_C\,, \label{grover_phi4}
\end{eqnarray}
where
$ | v_1 \rangle_C = (1,-1,-1,1)^T$,  $ | v_2 \rangle_C = (1, 1,0,0)^T$, $ | v_3 \rangle_C = (0,-1,1,0)^T$ and $ | v_4 \rangle_C = (-1,0,0,1)^T$. These eigenstates correspond to the eigenvalues $\alpha = \{-1, 1,1,1\}$, respectively. The addition denoted by $\oplus$  takes the boundary conditions into account: for reflecting boundary conditions (\textit{e. g.} carpet) the part of the states leaning over the boundary of the graph should be omitted (its amplitude is zero and the corresponding superposition state is normalized accordingly), and for periodic boundary conditions (\textit{e. g.} torus) the addition $\oplus$ corresponds to modulo operations with respect to the graph size.

Using these common eigenstates all p-attractors can be
constructed  by employing equation (\ref{pureattractors}). Performing the general analysis results, that the only non-p-attractor is the trivial one, which is proportional to identity. Thus, the total number of attractors is $(MN+M+N+1)^2 +
1$ for all carpets, and $(MN+1)^2+1$ for tori if $M$ or $N$ are odd.
However, in the latter case (\ref{grover_phi3}) and (\ref{grover_phi4}) are restricted by the periodic boundary conditions --- they cannot be used to construct attractors.
When $M$ and $N$ are both even in the case of tori, a single additional state from  (\ref{grover_phi3}) or (\ref{grover_phi4}) can be chosen as an additional common eigenstate. This results an attractor space with total number of attractors $(MN+2)^2+1$.

Analyzing the structure of the eigenstates reveals their connection with the effect of trapping.
The common eigenstates $| \phi_2 (s,t) \rangle$  have finite support. Consequently, these states
cannot be sensitive to boundary conditions, thus one can expect that they remain common
eigenstates even on an infinite system. Moreover, these states
are responsible for the trapping (localization) effect: An initially localized state overlapping with a $| \phi_2 (s,t)
\rangle$ state can always be found at its initial position with finite probability. The trapping effect for the percolation graph is illustrated in figure \ref{fig:grover}.  In addition, the family of pure localized eigenstates $| \phi_2 (s,t)
\rangle$ form a subspace which is free from the decoherence effects of the dynamical percolation. Such decoherence-free subspace might be quite useful, \textit{e. g.} serve as a quantum memory.

%%%%%%%%%%%%%%%%%%%%%%%%%%%%%
\section{Discussion}
%%%%%%%%%%%%%%%%%%%%%%%%%%%%%

Quantum walks by design can be employed to model physical processes. Ideally, the walk is described by a time stationary unitary process, thus its evolution is deterministic. However, imperfections of the physical system realizing the walk can enter the system, leading to decoherence and unexpected new effects. In the present paper we considered a special type of noise: graph with broken links. We gave a brief overview of the research field considering this effect, and also gave an extended presentation of our own results.

The system we study consists of graphs, where the links are dynamically broken. This type of perturbation is usually referred to as dynamical percolation. In our model, the evolution at a broken link is unitary, thus, the complete evolution is described by random unitary operations. The general method developed for the determination of asymptotics for RUO maps can be substantially simplified in the special case of quantum walks on dynamical percolation graphs, providing explicit solutions for a number of special cases. A further interesting refinement of the method consist of first searching for common eigenstates of the random unitaries, and then constructing the asymptotic solution from them. This pure state method makes it much easier to find the explicit form of the attractors. In order to complement our earlier results here we explicitly solved the problem of a general discrete time quantum walk on a one-dimensional percolation graph with arbitrary $SU(2)$ coin operator.

We found a new type of asymptotic solutions, the edge states \textit{i. e.} robust invariant pure quantum states exponentially localized at the boundaries of a linear one-dimensional graph (chain). Edge states enable, for example, asymptotic localization in the sense that an initially uniform position distribution will evolve to a highly peaked distribution at one of the edges of a chain. The coin operator crucially influences the form of the attractors, which allows for designing systems with a desired long time dynamics. 
 
In the two-dimensional case explicit solutions are known for some special coins. The coin space is much larger here ( $SU(4)$ ) and largely unexplored. We highlighted the Grover walk, as an example, where the effect of trapping (localization) survives the strong decoherence caused by the dynamically broken links. Behind the effect of localization one can identify a family of robust pure eigenstates with finite support, which form a decoherence-free subspace. Such a subspace might be useful in quantum information, \textit{e. g.} for information storage. 

In state of the art experiments \cite{Schreiber2011,Schreiber2012}, there is a considerable amount of dynamical control available, thus, there is hope, that the newly discovered effects might soon be demonstrated.

%%%%%%%%%%%%%%%%%%%%%%%%%%%%%
\section*{Acknowledgements}
%%%%%%%%%%%%%%%%%%%%%%%%%%%%%

We acknowledge support by GACR 13-33906S, RVO 68407700, the
Hungarian Scientific Research
Fund (OTKA) under Contract Nos. K83858, NN109651,
the Hungarian Academy of Sciences (Lend\"ulet Program,
LP2011-016).

%%Generated bibliography%%%

\end{document}